\documentclass[journal,onecolumn]{IEEEtran}

\usepackage[final]{graphicx}
\usepackage[reqno]{amsmath}
\usepackage{amssymb}
\usepackage{url}

\newfont{\bbb}{msbm10 scaled 500}

\newfont{\bb}{msbm10 scaled 1100}

\newcommand{\ZZ}{\mbox{\bb Z}}

\newcommand{\Prob}{\textrm{Pr}}

\newcommand{\Hm}{{\bf H}}
\newcommand{\Id}{{\bf I}}

\newcommand{\Xm}{{\bf X}}
\newcommand{\Ym}{{\bf Y}}
\newcommand{\Zm}{{\bf Z}}

\newcommand{\Ac}{{\cal A}}

\newcommand{\Cc}{{\cal C}}

\newcommand{\Ec}{{\cal E}}

\newcommand{\Hc}{{\cal H}}
\newcommand{\Ic}{{\cal I}}

\newcommand{\Kc}{{\cal K}}
\newcommand{\Lc}{{\cal L}}

\newcommand{\Nc}{{\cal N}}

\newcommand{\Sc}{{\cal S}}
\newcommand{\Tc}{{\cal T}}

\newcommand{\Wc}{{\cal W}}

\newtheorem{theorem}{Theorem}
\newtheorem{corollary}[theorem]{Corollary}
\newtheorem{lemma}[theorem]{Lemma}

\IEEEoverridecommandlockouts

\title{On Secrecy Capacity Scaling in\\ Wireless Networks}

\author{
\IEEEauthorblockN{O.~Ozan~Koyluoglu,~\IEEEmembership{Student~Member,~IEEE,}
C.~Emre~Koksal,~\IEEEmembership{Member,~IEEE,} and
Hesham~El~Gamal,~\IEEEmembership{Fellow,~IEEE}}
\thanks{This work is submitted to the IEEE Transactions on Information Theory.}
\thanks{The authors are with the Department of Electrical and Computer Engineering,
The Ohio State University,
Columbus, OH 43210, USA.
Email: \{koyluogo, koksal, helgamal\}@ece.osu.edu.}
\thanks{This work is partially supported by Los Alamos National Labs (LANL)
and by National Science Foundation (NSF).}
}

\begin{document}
\maketitle


\begin{abstract}
This work studies the achievable secure rate per source-destination pair in
wireless networks. First, a path loss model is considered, where the
legitimate and eavesdropper nodes are assumed to be placed according
to Poisson point processes with intensities $\lambda$ and $\lambda_e$,
respectively. It is shown that, as long as
$\lambda_e/\lambda=o\left((\log n)^{-2}\right)$,
almost all of the nodes achieve a perfectly secure rate of
$\Omega\left(\frac{1}{\sqrt{n}}\right)$ for the extended and
dense network models. Therefore, under these assumptions, securing the
network does not entail a loss in the per-node throughput.
The achievability argument is based on a novel multi-hop forwarding scheme
where randomization is added in every hop to ensure maximal ambiguity at
the eavesdropper(s). Secondly, an ergodic fading model with $n$
source-destination pairs and $n_e$ eavesdroppers is considered. Employing
the ergodic interference alignment scheme with an appropriate secrecy
pre-coding, each user is shown to achieve a constant positive secret
rate for sufficiently large $n$. Remarkably, the scheme does not require
eavesdropper CSI (only the statistical knowledge is assumed) and the secure
throughput per node increases as we add more legitimate users to the
network in this setting. Finally, the effect of eavesdropper collusion
on the performance of the proposed schemes is characterized.
\end{abstract}


\section{Introduction}
\label{sec:Introduction}

\subsection{Background}

In their seminal work~\cite{Gupta:The00} Gupta and Kumar have shown
that the randomly located nodes can achieve at most a rate that
scales like $\frac{1}{\sqrt{n}}$, as the number of nodes
$n\rightarrow \infty$, under an interference-limited channel model.
However, the proposed multi-hop scheme of~\cite{Gupta:The00} only
achieves a scaling of $\frac{1}{\sqrt{n\log n}}$ per node. This gap
was recently closed in~\cite{Franceschetti:Closing07}, where the
authors proposed a {\em highway} based multi-hop forwarding protocol
that achieves $\frac{1}{\sqrt{n}}$ rate per source-destination pair
in random networks. In this approach, a set of connected highways,
which span the network both horizontally and vertically, are
constructed. Then, each source-destination pair communicates via a
time-division strategy, where the source first transmits its message
to the closest horizontal highway. Then, the message is transported
in multi-hop fashion to the appropriate vertical highway, which
carries the message as close to the destination as possible.
Finally, the message is delivered to the destination node from the
vertical highway. The existence of highways, which satisfy certain
desirable properties, is established by borrowing tools from
percolation theory. Contrary to this multi-hop approach,
a single-hop scheme called as ergodic interference
alignment~\cite{Nazer:Ergodic09}
(see also~\cite{Cadambe:Interference08},~\cite{Maddah-Ali:Communication08})
is recently employed in~\cite{Niesen:Interference} and,
with arbitrary node placement and arbitrary traffic pattern,
the unicast and multicast capacity regions of dense networks
are characterized (up to a factor of $\log n$)
under the Gaussian fading channel model.
These line of works assumed an interference-limited channel model,
where the interference is considered as noise
(the focus of this work as well).
Contrary to this model,~\cite{Ozgur:Hierarchical07} considered Gaussian
fading channel model and proposed hierarchical cooperation schemes
that can increase the per-node rate. This approach is further
improved in the follow-up works (see,
e.g., \cite{Niesen:On09}, \cite{Ghaderi:Hierarchical09},
and references therein).

The broadcast nature of the wireless communication makes it susceptible
to eavesdropping. This motivates considering {\em secrecy}
as a quality of service (QoS) constraint that must be accounted
for in the network design. State of the art cryptographic approaches
can be broadly classified into public-key and private-key protocols.
Public-key cryptography assumes that the eavesdropper(s) has limited
computational power, whereas the decryption requires a significant
complexity without the knowledge of the key~\cite{DelfsAndKnebl:07}.
Private-key approaches, on the other hand,
assume that a random key is shared in private between the legitimate
transmitter and receiver. This key is used to secure the transmitted
information from potential eavesdropper(s). One of the earliest
examples of private-key cryptography is  the Vernam's one time
pad scheme~\cite{Vernam:OneTimePad26},
where the transmitter sends the XOR of the message bits
and key bits. The legitimate receiver can decode the messages by
XORing the shared key with the received sequence.
In~\cite{Shannon:Communication49}, Shannon showed that this scheme
achieves perfect secrecy {\bf if and only if} the two nodes share
a key of the same length as the message. The scaling laws of
wireless networks under the assumption of {\bf pre-distributed}
private keys was studied in~\cite{Bhandari:Secure08}. However,
it is important to note that, the key agreement step of the
cryptographic protocols is arguably the most challenging part and
this step becomes even more daunting as the network size grows.
Our work avoids the aforementioned limitations by adopting an
information theoretic framework for secrecy in wireless networks.
In particular, we assume the presence of eavesdropper(s) with
{\bf infinite computational power} and characterize the scaling
laws of the network secrecy capacity while {\bf relaxing the
idealistic assumption of pre-distributed keys}.

The notion of information theoretic secrecy was introduced by
Shannon to study secure communication over point-to-point noiseless
channels~\cite{Shannon:Communication49}. This line of work
was later extended by Wyner~\cite{Wyner:The75} to noisy channels.
Wyner's degraded wiretap channel assumes that the eavesdropper
channel is a degraded version of the one seen by the legitimate
receiver. Under this assumption, Wyner showed that the
advantage of the main channel over that of the eavesdropper,
in terms the lower noise level,
can be exploited to transmit secret bits using random binning codes.
This \emph{keyless secrecy} result was then extended to a more
general (broadcast) model in~\cite{Csiszar:Broadcast78} and to
the Gaussian setting in~\cite{Leung-Yan-Cheong:The78}. Recently,
there has been a renewed interest in wireless physical layer
security (see, e.g., Special Issue on Information
Theoretic Security, \emph{IEEE Trans. Inf. Theory}, June 2008
and references therein).
The secrecy in stochastic networks is studied in~\cite{Haenggi:The08},
where it is shown that even a small density of eavesdroppers
has a drastic impact on the connectivity of the secrecy graph.
Connectivity in stochastic networks with secrecy constraints
is also studied in~\cite{Pinto:Physical-layer08,Pinto:Wireless09},
where the node degree distribution is analyzed.
However, according to the best of our knowledge,
information theoretical analysis of secrecy capacity scaling in
large wireless networks has not been studied in the literature before.

\subsection{Contributions}

This paper considers wireless networks with secrecy constraints.
We study two different channel models: 1) Static path loss model, and
2) ergodic fading model. For the first model, we consider a stochastic
node placement on a square region, where the legitimate nodes and
eavesdroppers are distributed according to Poisson point processes
with intensity $\lambda$ and $\lambda_e$, respectively.
(For extended networks, the area of the region is $n$ and
$\lambda=1$; and, for dense networks, area of the region is
$1$ and $\lambda=n$.) The path loss is modeled with a power
loss exponent of $\alpha>2$. This model suits for the scenarios where
the channel gains are mostly determined by path losses.
In the second model, $n$ source-destination pairs and $n_e$
eavesdroppers are considered, where the gain of each link is
assumed to follow some fading process. (The assumptions on the
fading processes will be clear in the next section. Here,
we note that our model includes a large set of fading distributions.)
Arguably, this model suits for (dense) networks in which the inter
node distances have a negligible effect on the channel gains
compared to that of the underlying fading processes.

The results of this work can be summarized as follows.

1) For the path loss model, we construct a "highway backbone" similar
to~\cite{Franceschetti:Closing07}. However, in addition to the
interference constraint considered in~\cite{Franceschetti:Closing07},
our backbone construction and multi-hop forwarding strategy
are designed to ensure secrecy. More specifically, an edge can be used
in the highway if and only if there is a legitimate node within the
corresponding square of the edge and if there is no eavesdropper
within a certain \emph{secrecy zone} around the node.
We show that the network still percolates in this \emph{dependent}
edge model, and many highway paths can be constructed. Here,
in addition to the careful choice of the secrecy zone, our novel
multi-hop strategy, which enforces the usage of an {\em independent
randomization} at each hop, allows the legitimate nodes to create
an advantage over the eavesdroppers, which is, then, exploited to
transmit secure bits over the highways.  This way, we show that,
as long as $\lambda_e/\lambda=o\left((\log n)^{-2}\right)$,
almost all source-destination pairs achieve a secure rate
of $\Omega\left(\frac{1}{\sqrt{n}}\right)$ with high probability,
implying that the secrecy constraint does not entail a loss in the
per-node throughput (in terms of the scaling). (Note that
$\lambda=1$ for extended networks and $\lambda=n$ for dense
networks.) In these scenarios, the proposed scheme, which uses
independent randomization at the transmitter of each hop, is
the crucial step to obtain the results.

2) For the ergodic fading model,
employing the ergodic interference alignment scheme
(\cite{Nazer:Ergodic09},~\cite{Cadambe:Interference08},~\cite{Maddah-Ali:Communication08})
with an appropriate secrecy pre-coding we show that each
user can achieve secrecy. Here, the secrecy rate per user is shown
to be positive for most of the relevant fading distributions.
In particular, in the high SNR regime, the proposed scheme allows
each user to achieve a secure degrees of freedom of
$\eta=[\frac{1}{2}-\frac{1}{n}]^+$ even
with the absence of eavesdropper CSI. We observe that, per node
performance of users \emph{increase} as we add more legitimate
users in the network for this scenario compared to the
result obtained for the path loss model.

3) Finally, we focus on the eavesdropper collusion, where the
eavesdroppers are assumed to share their observations freely.
For the extended networks with the path loss model,
the same scaling result is shown to hold
for the colluding eavesdropper scenario when
$\lambda_e=O\left((\log n)^{-2(1+p)}\right)$ for any $p>0$.
For the ergodic fading model, extensions to many eavesdropper
collusion scenarios are discussed. In the extreme case, where all
the eavesdroppers collude, it is shown that the proposed scheme
allows each user to achieve a secure degrees of freedom of
$\eta=[\frac{1}{2}-\frac{n_e}{n}]^+$. We note that,
for the path loss model under the stated assumptions,
the eavesdropper collusion does not affect the performance of our
multi-hop scheme (in terms of scaling).
On the contrary, for the ergodic fading model, the eavesdropper
collusion has a clear effect on the achievable performance
of our ergodic interference alignment scheme.

\subsection{Organization}

The rest of this paper is organized as follows.
Section~\ref{sec:SystemModel} introduces the two network models
(path loss and ergodic fading models).
In Section~\ref{sec:PathLoss}, we consider
the path loss model and develop our novel
multi-hop secret encoding scheme. Section~\ref{sec:ErgodicFading}
focuses on the ergodic fading scenario and proposes ergodic interference
alignment scheme for security applications.
In Section~\ref{sec:Colluding}, we focus on the colluding eavesdropper
scenarios. Concluding remarks are given in Section~\ref{sec:Conclusion},
and, to enhance the flow of the paper, some of technical lemmas and proofs
are relegated to the Appendix.


\section{Network Models}
\label{sec:SystemModel}

The set of legitimate nodes is denoted by $\Lc$, whereas the
set of eavesdroppers is represented by $\Ec$. During time slot $t$,
the set of transmitting nodes are denoted by $\Tc(t)\subset\Lc$, where
each transmitting user $i\in\Tc(t)$ transmits the signal $X_i(t)$.
The received signals at receiving node $j\in\Lc-\Tc(t)$ and
at eavesdropper $e\in\Ec$ are denoted by $Y_j(t)$ and
$Y_e(t)$, respectively:
\begin{eqnarray}
Y_j(t)&=&\sum_{i\in\Tc(t)} h_{i,j}(t) X_i(t) + Z_j(t)\label{eq:Model1}\\
Y_e(t)&=&\sum_{i\in\Tc(t)} h_{i,e}(t) X_i(t) + Z_e(t)\label{eq:Model2},
\end{eqnarray}
where receivers are impaired by zero-mean circularly symmetric
complex Gaussian noises with variance $N_0$.
We denote this distribution by $\Cc\Nc\left(0,N_0\right)$.
Assuming that each transmitter is active over $N$ channel uses,
the average power constraint on channel inputs at each transmitter
is given by $\frac{1}{N}\sum\limits_{t=1}^N |X_i(t)|^2 \leq P$.
Note that, for i.i.d.~$\Cc\Nc(0,P)$ input distribution,
$\textrm{SNR} \triangleq \frac{P}{N_o}$ is the signal-to-noise ratio
per complex symbol.

\subsection{Static Path Loss Model with Stochastic Node Distribution}
\label{sec:SystemModel1}

In the path loss model we consider, the signal power decays with the
distance $d$ as $d^{-\alpha}$ for some $\alpha > 2$; and
the distance between node $i$ and node $j$ is denoted by $d_{ij}$.
The path loss is modeled in \eqref{eq:Model1} and
\eqref{eq:Model2} with
\begin{eqnarray}
h_{i,j}(t) = \sqrt{d_{i,j}^{-\alpha}},\quad
h_{i,e}(t) = \sqrt{d_{i,e}^{-\alpha}}.
\end{eqnarray}
The set of all observations at eavesdropper
$e$ is denoted by $\Ym_e\triangleq \{Y_e(t), \forall t\}$.

The extended network model is a square of side-length $\sqrt{n}$ (the
area of the region is $n$). The legitimate nodes and eavesdroppers are
assumed to be placed randomly according to Poisson point processes of
intensity $\lambda=1$ and $\lambda_e$, respectively.
The transmitters are assumed to know {\em a-priori} whether
there is any eavesdropper within some neighborhood or not
(the size of the neighborhood will be clear in later parts
of the text). We are aware of the idealistic nature of this
assumption, but believe that it allows for extracting valuable
insights in the problem.
To analyze the worst case scenario from a security
perspective, the legitimate receivers are assumed to
consider interference as noise, whereas no such assumption is made
on the eavesdroppers, all of which are assumed to be informed
with the network topology perfectly.

Now, consider any random source-destination pair, where the source
$s$ wishes to transmit the message $W_{s,d}$ securely to the
intended destination $d$. In our multi-hop strategy, each
transmission consists of $N$ channel uses per hop. We say that the secret
rate of $R$ is achievable for almost all the source-destination
pairs ($s,d$), if
\begin{itemize}
  \item The error probability of decoding the intended message
  at the intended receiver can be made arbitrarily small as $N\to \infty$, and
  \item The information leakage rate
  associated with the transmissions of the message over the
  entire path, i.e., $\frac{I(W_{s,d};\Ym_e)}{N}$, can be
  made arbitrarily small $\forall e\in \Ec$ as $N \to \infty$,
\end{itemize}
for almost all ($s,d$).

If there are $H$ hops carrying the
message $W_{s,d}$, one only needs to consider the associated channel
observations at the eavesdropper when evaluating our security constraint.
Hence, our second condition is satisfied if
$\frac{I(W_{s,d};\Ym_{e}(1),\cdots,\Ym_{e}(H))}{N}$
can be made arbitrarily small for sufficiently large block lengths,
where $\Ym_{e}(h)$ denotes the length-$N$ channel output vector
at eavesdropper $e\in\Ec$ during hop $h$~\footnote{
We note that the length of the observation vector $\Ym_e$
regarding message $W_{s,d}$ is $NH$ for $H$ hops and
$N$ channel uses per hop. Therefore,
to analyze the mutual information leakage rate per channel use
one might be tempted to use $\frac{I(W_{s,d};\Ym_{e}(1),\cdots,\Ym_{e}(H))}{NH}$
in the secrecy constraint. However, as $H$ hops carry the same message
$W_{s,d}$, the overall information accumulation at the eavesdropper
might be large even if $\frac{I(W_{s,d};\Ym_{e}(1),\cdots,\Ym_{e}(H))}{NH}$
is made arbitrarily small.
}.

To derive our asymptotic scaling results, we use the following
probabilistic version of Landau's notation.
We say $f(n)=O(g(n))$ w.h.p., if there exists a constant
$k$ such that
$$\lim\limits_{n\to\infty} \Prob
\left\{ f(n) \leq k g(n) \right\} = 1.$$
We also say that $f(n)=\Omega(g(n))$ w.h.p., if
w.h.p. $g(n) = O(f(n))$. We denote $f(n)=\Theta(g(n))$,
if $f(n)=O(g(n))$ and $f(n)=\Omega(g(n))$.
Lastly, we say $f(n)=o(g(n))$,
if $\frac{f(n)}{g(n)}\to 0$, as $n\to\infty$.

We also analyze a dense networks with the path loss model
and stochastic node distribution similar to above, where
we assume that the network is deployed on a square
region of unit area. In this case, we assume that
the legitimate nodes have an intensity of $\lambda=n$.

\subsection{Ergodic Fading Model}
\label{sec:SystemModel2}

Fading process for the link from $i$ to $k$, denoted
by $h_{i,k}(t)$, is assumed to be
drawn i.i.d. across time according to some ergodic fading process.
The ergodic fading is modeled in \eqref{eq:Model1} and
\eqref{eq:Model2} with the following two assumptions:
\begin{itemize}
  \item The channel gains for the legitimate users,
$h_{i,j}$, are assumed to be drawn from
independent distributions (for each $i,j\in\Kc$) that
are symmetric around zero (that is
$\Prob\{h_{i,j}=h\}=\Prob\{h_{i,j}=-h\}$); and
  \item The fading process for eavesdropper $e\in\Ec$, i.e., $h_{i,e}$,
is assumed to be drawn independently from the same distribution
$\forall i\in\Kc$.
\end{itemize}
Note that, as we assume a certain distribution for any given
transmitter-receiver pair, the location of the nodes are not
relevant in this model. In addition, the second assumption
on the fading processes ensures that each eavesdropper
has statistically the same channel to each transmitter.

We denote $\Ym_e\triangleq \{Y_e(1),\cdots, Y_e(N)\}$,
$\Hm(t)\triangleq \{h_{i,j}(t), \forall i,j\in\Kc\}$,
$\Hm \triangleq \{\Hm(1),\cdots, \Hm(N)\}$,
$\Hm_e(t)\triangleq \{h_{i,e}(t), \forall i\in\Kc, \forall e\in\Ec\}$,
and $\Hm_e \triangleq \{\Hm_e(1),\cdots, \Hm_e(N)\}$.
Here, $\Hm$ is assumed to be known at legitimate users,
whereas eavesdroppers are assumed to know both $\Hm$ and $\Hm_e$.

We assume that each transmitter in the network has an arbitrary
and distinct receiver and the set of legitimate nodes, i.e.,
$\Lc$, consists of $n$ source-destination pairs.
For notational convenience, we enumerate each transmitter-receiver
pair using an element of $\Kc=\{1,\cdots,n\}$, and denote the
channel gain process associated with transmitter-receiver
pair $i$ with $h_{i,i}(t)$.
In this model, transmitter-receiver pair $i\in\Kc$ tries to
communicate a secret message $W_i\in\Wc_i$. Denoting
the decoding error at the receiver by $P_{e,i}$,
we say that the secret rate $R_i$ is achievable, if
for any $\epsilon>0$,
1) $|\Wc_i|\geq 2^{NR_i}$, 2) $P_{e,i}\leq \epsilon$,
and 3) $\frac{1}{N}I(W_i;\Ym_e,\Hm,\Hm_e) \leq \epsilon$,
$\forall e\in\Ec$,
for sufficiently large $N$.
We finally say that the symmetric secure degrees of freedom (DoF) (per
orthogonal dimension) of $\eta$ is achievable, if
the rate $R_i$ is achievable for pair $i\in\Kc$ and
\begin{eqnarray}
\eta \leq \lim
\limits_{\textrm{SNR} \to\infty}
\frac{R_i}{ \log(\textrm{SNR})}, \forall i\in\Kc.
\end{eqnarray}


\section{The Path Loss Model}
\label{sec:PathLoss}

In this section, we first focus on extended networks with a path
loss model ($\alpha>2$) and stochastic node distribution
(Poisson point processes) as detailed in Section~\ref{sec:SystemModel1}.
Our achievability argument is divided into the following four key steps:
\begin{enumerate}
\item Lemma~\ref{thm:Lemma1:SecureRate} uses the idea of {\bf
secrecy zone} to guarantee the secrecy of the communication over a single hop.
\item In Lemma~\ref{thm:Lemma2:MultihopSecurity}, we introduce
our novel multi-hop forwarding strategy which uses independent
randomization signal in each hop. This strategy is shown to allow
for hiding the information from an eavesdropper which listens to
the transmissions over {\bf all} hops.
\item Using tools from percolation theory, we show the
existence of a sufficient number of horizontal and vertical
highways in Lemma~\ref{thm:Lemma3:Percolation}, and
we characterize the rate assigned to each node on the highway
in Lemma~\ref{thm:Lemma4:HighwayRate}.
\item The accessibility of highways for {\bf almost all} the
nodes in the networks with the appropriate rates is established
in Lemma~\ref{thm:Lemma5:AccessRate}.
\end{enumerate}
Our main result, i.e., Theorem~\ref{thm:Theorem1}, is
then proved by combining the aforementioned steps
with a multi-hop routing scheme (Fig.~$1$).

We partition the network area into squares of constant side length
$c$. We further divide the area into larger squares of side $f_tdc$,
each of which contains $(f_td)^2$ small squares. These small squares
take turn over a Time-Division-Multiple-Access (TDMA) frame of size
$(f_td)^2$ slots. In each slot, a transmitter within each active small
square can transmit to a receiver that is located at most $d$ squares
away as illustrated in Fig.~$2$. On the same figure, we also show the
secrecy zone, around a transmitting square, consisting of squares
that are at most $f_e d$ squares away. Our first result establishes
an achievable {\bf secure} rate per {\bf a single hop}, active over
$N$ channel uses, under the assumption of a single eavesdropper on
the boundary of the secrecy zone.

\begin{lemma}[Secure Rate per Hop]\label{thm:Lemma1:SecureRate}
In a communication scenario depicted in Fig.~$2$, the secure rate,
simultaneously achievable between any active transmitter-receiver pair is:
\begin{equation}\label{eq:Lemma1eq8}
R_{TR} = \frac{1}{(f_td)^2}
\left[\log(1 + \underline{\textrm{SNR}_{TR}})
- \log(1 + \overline{\textrm{SNR}_{e^*}})\right],
\end{equation}
where
\begin{eqnarray}
\underline{\textrm{SNR}_{TR}}
&\triangleq& \frac{P (d+1)^{-\alpha}c^{-\alpha} (\sqrt{2})^{-\alpha}}
{N_o + P 8 (f_t)^{-\alpha}d^{-\alpha}c^{-\alpha} S(\alpha)},\label{eq:Lemma1eq5}\\
S(\alpha) &\triangleq& \sum\limits_{i=1}^{\infty} i(i-0.5)^{-\alpha},\\
\overline{\textrm{SNR}_{e^*}}
&\triangleq& \frac{P (f_e)^{-\alpha}d^{-\alpha}c^{-\alpha}}{N_o},\label{eq:Lemma1eq6}\\
f_t &\geq& \frac{2(d+1)}{d},
\end{eqnarray}
and
\begin{equation}\label{eq:Lemma1eq7}
\frac{(d+1)^\alpha (\sqrt{2})^{\alpha}}{ (d)^\alpha }
\left[ 1 + \frac{P}{N_o} 8
(f_t)^{-\alpha}d^{-\alpha}c^{-\alpha} S(\alpha) \right] < (f_e)^\alpha.
\end{equation}
Here, secrecy is guaranteed assuming the presence of an
eavesdropper on the boundary of the secrecy zone.
\end{lemma}
\begin{IEEEproof}
In Fig.~$2$, consider that one node per filled square is transmitting.
Assuming that there is a transmission from every such square,
we denote the interference set seen by our designated legitimate
receiver as $\Ic$. Since the legitimate receivers simply
consider other transmissions as noise in our model, we
obtain the following SNR at the legitimate receiver.
\begin{equation}\label{eq:Lemma1eq1}
\textrm{SNR}_{TR} =
\frac{P d_{TR}^{-\alpha}}{N_o
+ \sum\limits_{i\in\Ic} P d_{iR}^{-\alpha}},
\end{equation}
where the distance between the transmitter and receiver is
denoted as $d_{TR}$ and that between interferer $i\in\Ic$
and our receiver is denoted by $d_{iR}$.

We now consider an eavesdropper $e\in\Ec$ listening to the
transmission and upper bound its received SNR by the
following.
\begin{equation}\label{eq:Lemma1eq2}
\textrm{SNR}_{e}
\leq \frac{P d_{Te}^{-\alpha}}{N_o},
\end{equation}
where the distance between the transmitter and the eavesdropper $e$
is denoted by $d_{Te}$. Here, the upper bound follows by eliminating
the interference at the eavesdropper. The construction in Fig.~$2$
allows for showing that
\begin{eqnarray}
d_{TR} &\leq& (d+1)c\sqrt{2}, \label{eq:Lemma1eq3-1}\\
d_{Te} &\geq& f_e d c, \label{eq:Lemma1eq3-2}
\end{eqnarray}
and
\begin{eqnarray}\label{eq:Lemma1eq4}
\sum\limits_{i\in\Ic} d_{iR}^{-\alpha}
&=& \sum\limits_{i=1}^{\infty}
8i(if_td-(d+1))^{-\alpha}c^{-\alpha} \nonumber\\
&\overset{(a)}{\leq}& 8(f_tdc)^{-\alpha} \sum\limits_{i=1}^{\infty}
i(i-0.5)^{-\alpha} \nonumber\\
&=& 8(f_tdc)^{-\alpha} S(\alpha),
\end{eqnarray}
where $(a)$ follows by choosing
\begin{equation}
f_t d \geq 2(d+1),
\end{equation}
and the last equality follows
by defining
\begin{equation}
S(\alpha) \triangleq \sum\limits_{i=1}^{\infty} i(i-0.5)^{-\alpha},
\end{equation}
which converges to some finite value as $\alpha > 2$.

Using~\eqref{eq:Lemma1eq3-1},~\eqref{eq:Lemma1eq3-2},~\eqref{eq:Lemma1eq4}
in~\eqref{eq:Lemma1eq1} and~\eqref{eq:Lemma1eq2},
we obtain the followings.
\begin{equation}
\textrm{SNR}_{TR} \geq \underline{\textrm{SNR}_{TR}}
\triangleq \frac{P (d+1)^{-\alpha}c^{-\alpha} (\sqrt{2})^{-\alpha}}
{N_o + P 8 (f_t)^{-\alpha}d^{-\alpha}c^{-\alpha} S(\alpha)},
\end{equation}
and
\begin{equation}
\textrm{SNR}_{e} \leq \overline{\textrm{SNR}_{e^*}}
\triangleq \frac{P (f_e)^{-\alpha}d^{-\alpha}c^{-\alpha}}{N_o}.
\end{equation}

Hence, $\textrm{SNR}_{TR} > \textrm{SNR}_{e}$ for every
eavesdropper $e$,
once we choose $f_e$ such that
\begin{equation}
\frac{(d+1)^\alpha (\sqrt{2})^{\alpha}}{ (d)^\alpha }
\left[ 1 + \frac{P}{N_o} 8
(f_t)^{-\alpha}d^{-\alpha}c^{-\alpha} S(\alpha) \right] < (f_e)^\alpha.
\end{equation}

We then construct the secrecy codebook at the transmitter
by considering an eavesdropper that observes
the signals of the transmission of {\bf this hop only}
with an SNR of $\overline{\textrm{SNR}_{e^*}}$. Based on
the Gaussian wiretap channel capacity~\cite{Leung-Yan-Cheong:The78},
one can easily show that the following {\bf perfectly secure}
rate is achievable
\begin{equation}
R_{TR} = \frac{1}{(f_td)^2}
\left[\log(1 + \underline{\textrm{SNR}_{TR}})
- \log(1 + \overline{\textrm{SNR}_{e^*}})\right],
\end{equation}
where the $(f_td)^2$ term is due to time-division described above.
\end{IEEEproof}

Next we introduce our novel multi-hop {\em randomization} strategy
which ensures secrecy over the \emph{entire
path}, from a source to a destination node,
at \emph{every} eavesdropper observing {\em all} transmissions.

\begin{lemma}[Securing a Multi-Hop Path]\label{thm:Lemma2:MultihopSecurity}
Securing each hop from an eavesdropper that is located on the
boundary of the secrecy zone is sufficient to ensure secrecy from
any eavesdropper which listens the transmissions from all the hops
and lie outside the secrecy zones of transmitters of hops.
\end{lemma}

\begin{IEEEproof}
We consider a source $s$, a destination $d$, and an eavesdropper
$e$ in the network. Without loss of generality, we assume that
the multi-hop scheme uses $H$ hops to route the message. We design
the secrecy codebook at each transmitter according to highest possible
eavesdropper SNR assumption for each hop. In our multi-hop routing
scenario, each code of the ensemble at the transmitter of hop $i$
generates $2^{N(R_i+R^x_i-\frac{\epsilon_1}{H})}$ codewords each
entry with i.i.d.~$\Cc\Nc(0,P)$, for some $\epsilon_1>0$, and distributes
them into $2^{NR_i}$ bins. Each codeword is, therefore, represented
with the tuple $(w_{s,d},w^x_i)$, where $w_{s,d}$ is the bin index
(secret message) and $w^x_i$ is the codeword index (randomization message).
To transmit the message $w_{s,d}$, the encoder of transmitter $i$
will randomly choose a codeword within the bin $w_{s,d}$
according to a uniform distribution. This codeword, i.e.,
$\Xm_i(w_{s,d},w^x_i)$, is sent from transmitter $i$. It is clear now that
each transmitter on the path adds
\emph{independent} randomness, i.e., the codeword index $w^x_i$
is independent of $w^x_j$ for $i\neq j$.

We consider an eavesdropper at the boundary of the secrecy
zone around the transmitter of the hop $i$,
and denote it by $e^*_i$. We subtract all the interference
seen by this virtual node and denote its observations
for hop $i$ as $\Ym_{e^*_i}$. Omitting the indices $(w_{s,d},w^x_i)$, for simplicity,
we denote the symbols transmitted from the transmitter $i$ as $\Xm_i$; and set
$R^x_i=I(X_i;Y_{e^*_i})=\log\left(1+ \overline{\textrm{SNR}_{e^*_i}}\right)$.
(Note that this is the rate loss in~\eqref{eq:Lemma1eq8}.)
We continue as below.
\begin{eqnarray}
I(W_{s,d};\Ym_e)  &=&  I(W_{s,d};\Ym_{e}(1),\cdots,\Ym_{e}(H)) \nonumber\\
& \overset{(a)}{\leq} & I(W_{s,d};\Ym_{e^*_1},\cdots,\Ym_{e^*_H}) \nonumber\\
& = & I(W_{s,d},W^x_1,\cdots,W^x_H;\Ym_{e^*_1},\cdots,\Ym_{e^*_H})
- I(W^x_1,\cdots,W^x_H;\Ym_{e^*_1},\cdots,\Ym_{e^*_H}|W_{s,d}) \nonumber\\
& \overset{(b)}{\leq} & I(\Xm_1,\cdots,\Xm_H;\Ym_{e^*_1},\cdots,\Ym_{e^*_H})
- H(W^x_1,\cdots,W^x_H|W_{s,d})
+ H(W^x_1,\cdots,W^x_H|
\Ym_{e^*_1},\cdots,\Ym_{e^*_H},W_{s,d}) \nonumber\\
& \overset{(c)}{=} & \sum_{i=1}^H
I(\Xm_1,\cdots,\Xm_H;\Ym_{e^*_i}|\Ym_{e^*_1},\cdots,\Ym_{e^*_{i-1}})
- H(W^x_1,\cdots,W^x_H)\nonumber\\
&& {+}\: \sum_{i=1}^H H(W^x_i|W_{s,d},\Ym_{e^*_1},\cdots,\Ym_{e^*_H},
W^x_1,\cdots,W^x_{i-1}) \nonumber\\
& = & \sum_{i=1}^H \bigg[ I(\Xm_i;\Ym_{e^*_i}|\Ym_{e^*_1},\cdots,\Ym_{e^*_{i-1}})
+ I(\Xm_1,\cdots,\Xm_{i-1},\Xm_{i+1},\cdots,\Xm_H;\Ym_{e^*_i}|
\Ym_{e^*_1},\cdots,\Ym_{e^*_{i-1}},\Xm_i) \nonumber\\
&& {-}\: N R^x_i + N \frac{\epsilon_1}{H}
+ H(W^x_i|\Ym_{e^*_i},W_{s,d}) \bigg] \nonumber\\
& \overset{(d)}{\leq} & \sum_{i=1}^H \bigg[
H(\Ym_{e^*_i}|\Ym_{e^*_1},\cdots,\Ym_{e^*_{i-1}})
- H(\Ym_{e^*_i}|\Ym_{e^*_1},\cdots,\Ym_{e^*_{i-1}},\Xm_i)
- N R_i^x
+ N \frac{\epsilon_1+\epsilon_2}{H} \bigg] \nonumber\\
& \overset{(e)}{\leq} & \sum_{i=1}^H \bigg[ H(\Ym_{e^*_i})
- H(\Ym_{e^*_i}|\Xm_i)
- N R_i^x
+ N \frac{\epsilon_1+\epsilon_2}{H} \bigg] \nonumber\\
& = & \sum_{i=1}^H \bigg[ I(\Xm_i;\Ym_{e^*_i}) - N R_i^x
+ N \frac{\epsilon_1+\epsilon_2}{H} \bigg] \nonumber\\
& \overset{(f)}{\leq} & \sum_{i=1}^H \bigg[ NI(X_i;Y_{e^*_i}) - N R_i^x
+ N \frac{\epsilon_1+\epsilon_2}{H} \bigg] \nonumber\\
&=& N (\epsilon_1+\epsilon_2) \nonumber,
\end{eqnarray}
where (a) is due to the fact that
$\Ym_{e^*_i}$ is an enhanced set of observations
compared to that of $\Ym_{e}(i)$, (b) is due to the
data processing inequality and the Markov chain
$\{W_{s,d},W^x_1,\cdots,W^x_H\}
\to \{\Xm_1,\cdots,\Xm_H\}
\to \{\Ym_{e^*_1},\cdots,\Ym_{e^*_H}\}$,
(c) follows since
$W_{s,d}$ and $W_i^x$ are independent, (d) is
due to fact that the second term in the sum is zero and due to
Fano's inequality (as we choose $R^x_{i}\leq I(X_i;Y_{e^*_i})$,
the binning codebook construction allows for decoding randomization
message at the eavesdropper given the bin index for almost all codebooks
in the ensemble): We define the decoding error probability as
$P_{e,e^*_i}\triangleq \Prob\{\hat{W}_i^x \neq W_i^x\}$, where
$\hat{W}_i^x$ is the estimate of the randomization message $W_i^x$
given ($\Ym_{e^*_i},W_{s,d}$), and bound
\begin{eqnarray}\label{eq:Lemma2}
H(W^x_i|\Ym_{e^*_i},W_{s,d}) \leq N \left(
\frac{H(P_{e,e^*_i})}{N} + P_{e,e^*_i} R_i^x
\right) \leq N \frac{\epsilon_2}{H}
\end{eqnarray}
with some $\epsilon_2 \to 0$
as $N \to \infty$, (e) follows by the fact that conditioning does not
increase the entropy and the observation that
$H(\Ym_{e^*_i}|\Ym_{e^*_1},\cdots,\Ym_{e^*_{i-1}},\Xm_i)
=H(\Ym_{e^*_i}|\Xm_i)$, and (f) is due to the
fact that $I(\Xm_i;\Ym_{e^*_i})=\sum\limits_{t=1}^N
I(\Xm_i;Y_{e^*_i} (t)|Y_{e^*_i} (1),\cdots, Y_{e^*_i} (t-1))
\leq \sum\limits_{t=1}^N
H(Y_{e^*_i}(t)) - H(Y_{e^*_i}(t)|X_i(t))
= NI(X_i;Y_{e^*_i})$.

After setting, $\epsilon=\epsilon_1+\epsilon_2$, we
obtain our result: For any given $\epsilon>0$,
$\frac{I(W_{s,d};\Ym_e)}{N} < \epsilon$ as $N \to \infty$.
\end{IEEEproof}

Note that, the number of hops scale as $H=O(\sqrt{n})$
and in~\eqref{eq:Lemma2} we have $P_{e,e^*_i}$ decays
exponentially in $N$. Thus, we can say that the multi-hop
transmissions require larger block lengths, as $n$ gets large,
to assure secrecy with this scheme.

The following result uses tools from percolation theory to
establish the existence of a sufficient number of
{\bf secure highways} in our network.

\begin{lemma}[Secure Highways]\label{thm:Lemma3:Percolation}
There exist a sufficient number of \emph{secure} vertical and
horizontal highways such that, as $n\to\infty$,
each secure highway is required to serve
$O(\sqrt{n})$ nodes and an entry (exit) point has
w.h.p. a distance of at most $\kappa'\log n$
away from each source (respectively, destination)
for some finite constant $\kappa'>0$, if $c\geq c_0$
for some finite constant $c_0>0$ and $\lambda_e\to 0$.
\end{lemma}
\begin{IEEEproof}
We first describe the notion of secure highway and the
percolation model we use in the proof.
We note that most of this percolation based construction
is developed in~\cite{Franceschetti:Closing07,FranceschettiAndMeester:07}
and here we generalize it for secrecy. We say that each square is "open"
if the square has at least one legitimate node and there
are no eavesdroppers in the secrecy zone around the square.
We denote the probability of having at least one legitimate node in
a square by $p$. It is evident that
$$p=1-e^{-c^2},$$
and hence, $p$ can be made arbitrarily close to $1$ by increasing
$c$. For any given transmitting square, we denote the probability of
having an eavesdropper-free secrecy zone by $q$. The number of
eavesdroppers within a secrecy zone is a Poisson random variable
with parameter $\lambda_e(2f_e d + 1)^2 c^2$, and hence,
$$q = e^{-\lambda_e(2f_e d + 1)^2 c^2}.$$
Thus, $q$ gets arbitrarily close to $1$, as
$n\to\infty$, since $\lambda_e\to0$ with $n$ ($f_e$,
$d$, and $c$ are some finite numbers for the highway construction).

We then map this model to a discrete edge-percolation model
(a.k.a. bond percolation on the random square grid~\cite{Grimmett:99})
by drawing horizontal and vertical edges over the open squares, where
an edge is called open if the corresponding square is open (see Fig.~$3$).
We are interested in characterizing (horizontal and vertical) open paths
that span the network area. Such open paths are our
\emph{horizontal and vertical highways}. We only focus on horizontal
highways for the rest of the section as the results hold, due to
symmetry, for the vertical highways. We remark that, in our model,
the status of edges are not statistically independent due to the
presence of associated secrecy zones that intersect for successive
squares. Notice that the status of two edges would be independent if
their secrecy zones did not intersect, which happens if there were
at least $2f_ed$ squares between two edges. Therefore, this
dependent scenario is referred to as finite-dependent model, as
$f_e$ and $d$ are some finite numbers. Due to
Lemma~\ref{thm:Lemma6:DependentPercolation}, given in
Appendix~\ref{sec:Appendix1},
this dependent model \emph{stochastically dominates} an independent
model, in which edges are independently open with probability $p'$,
where $p'$ can be made arbitrarily high if $pq$ can be made
arbitrarily high. This independent scenario can be constructed by
following the steps provided in~\cite{Liggett:Domination97}.
Therefore, after proving the percolation of the network with some
desirable properties under the independence assumption, the network
will also percolate with the same properties under the finite
dependence model as both $p$ and $q$ can be made sufficiently large.

Using the independent edge model, applying
Lemma~\ref{thm:Lemma9:Percolation}, given in Appendix~\ref{sec:Appendix1},
with edge openness probability of $p'$, and noting the fact that
$m=\frac{\sqrt{n}}{c\sqrt{2}}$ (Fig.~$3$), we obtain the
following: There are w.h.p. $\Omega(\sqrt{n})$ horizontal paths,
which, for any given $\kappa > 0$, can be grouped into disjoint sets
of $\lceil \delta \log n \rceil$ highways that span a rectangle area
of size $(\kappa \log n - \epsilon) \times  \sqrt{n}$, for some
$\delta>0$, and some $\epsilon\to 0 $ as $n\to\infty$
if $p'$ is high enough. Then, the network area is sliced into slabs of
side length $w$, chosen so that the number of slabs match with the number
of highways in each rectangle. Then, each source (destination) in the $i$th
horizontal (vertical) slab will access the corresponding highway (Fig.~$4$).
This way, each highway is required to serve at most $2w\sqrt{n}$ nodes and
an entry (exit) point has w.h.p. a distance of at most $\kappa'\log n$
away from each source (respectively, destination) for some
finite constant $\kappa'>0$. The former claim follows by an
application of Chernoff bound, given in Lemma~\ref{thm:Lemma7:Chernoff},
and union bound (see~\cite[Lemma 2]{Franceschetti:Closing07}
or~\cite[Lemma 5.3.5]{FranceschettiAndMeester:07} for details)
and the latter incorporates the negligible horizontal distance
(at most $c\sqrt{2}$) in addition to the vertical distance,
which scales as $\kappa \log n$.
Finally, due to the statistical domination argument given above,
these percolation results will also hold for our finite-dependent
model, as $pq$ can be made arbitrarily large as $n\to\infty$.
Formally,
$\exists c_0\in(0,\infty)$ such that, for any
$c\geq c_0$, $pq$ can be made sufficiently high
if $\lambda_e\to 0$ as $n\to\infty$. This translates to high
enough $p'$ by Lemma~\ref{thm:Lemma6:DependentPercolation},
which shows that the dependent model has the property given
in Lemma~\ref{thm:Lemma9:Percolation} as well.
\end{IEEEproof}

With the following lemma we conclude the
discussion of highways.

\begin{lemma}[Rate per Node on the Highways]\label{thm:Lemma4:HighwayRate}
Each node on the constructed highways can transmit to
their next hop at a constant secure rate. Furthermore,
the number of nodes each highway serves is $O(\sqrt{n})$,
and therefore each highway can w.h.p. carry a per-node
secure throughput of $\Omega\left(\frac{1}{\sqrt{n}}\right)$.
\end{lemma}

\begin{IEEEproof}
The highways are constructed such that there is at least
one legitimate node per square and there
are no eavesdroppers within the secrecy zone around the
squares of the highway. We choose one legitimate node per
square as a member of the highway, and compute the rate
that can be achieved with the multi-hop strategy.
From Lemma~\ref{thm:Lemma1:SecureRate} (with $d=1$)
and Lemma~\ref{thm:Lemma2:MultihopSecurity}, one can
see that highways can carry data \emph{securely}
with a \emph{constant positive rate}.
As each highway carries the data for $O(\sqrt{n})$ nodes
due to Lemma~\ref{thm:Lemma3:Percolation},
the achievable rate per node on highways is
$\Omega\left(\frac{1}{\sqrt{n}}\right)$.
\end{IEEEproof}

Our final
step is to show that
almost all the nodes can access the highways simultaneously
with high probability with a rate scaling higher than
$\Omega\left(\frac{1}{\sqrt{n}}\right)$.

\begin{lemma}[Access Rate to Highways]\label{thm:Lemma5:AccessRate}
Almost all source (destination) nodes can w.h.p.
simultaneously transmit (receive)
their messages to (from)
highways with a secure rate of
$\Omega\left((\log n)^{-3-\alpha}\right)$, if
$\lambda_e=o\left((\log n)^{-2}\right)$.
\end{lemma}

\begin{IEEEproof}
To calculate the rate of each node transmitting to
the closest horizontal highway, we follow the same procedure
given in the proof of Lemma~\ref{thm:Lemma4:HighwayRate}.
However, this time we choose $d=\kappa'' \log n$
in Lemma~\ref{thm:Lemma1:SecureRate} for some finite $\kappa''>0$,
as the nodes within each transmitting squares need to transmit to a
receiver at a distance of at most $\kappa'' \log n$ squares
away (due to Lemma~\ref{thm:Lemma3:Percolation}).
(Here, we can choose smallest number $\kappa''\geq\frac{\kappa'}{c}$
making $\kappa'' \log n$ integer.)
In addition, compared to Lemma~\ref{thm:Lemma4:HighwayRate},
where only one node per square is transmitting, here
all legitimate nodes within small squares should access the
highways w.h.p., which is accomplished with a TDMA scheme.

As $d = \kappa''\log n \to\infty$, we see from
\eqref{eq:Lemma1eq5}, \eqref{eq:Lemma1eq6}, \eqref{eq:Lemma1eq8}
that a per-node rate of $\Omega\left( (\log n)^{-2-\alpha}\right)$
is achievable. Note that, to satisfy~\eqref{eq:Lemma1eq7} and
thus~\eqref{eq:Lemma1eq8}, any choice of $f_e > \sqrt{2}$ suffices
as $n\to\infty$. However, for this case, due to time division
between nodes within squares this rate needs to be further modified.
Again applying the Chernoff bound (Lemma~\ref{thm:Lemma7:Chernoff})
and the union bound one can show that there are w.h.p.
$O(\log n)$ legitimate nodes in each square
(see~\cite[Lemma 1]{Franceschetti:Closing07}
or~\cite[Lemma 5.3.4]{FranceschettiAndMeester:07} for details).
Therefore, w.h.p. the secure rate
$\Omega\left( (\log n)^{-3-\alpha}\right)$ is achievable
to the associated highway from a source node, if there is
{\bf no eavesdropper} in the associated secrecy zone.
Next, we show that this will happen with a very high
probability if $\lambda_e=o\left((\log n)^{-2}\right)$
asymptotically (as $n\to\infty$).

From Fig.~$2$, it is clear that the presence of
an eavesdropper eliminates the possibility of secure
access to a highway from a region of area
$A=(2f_ed+1)^2c^2$. We denote the total number of
eavesdroppers in the network as $|\Ec|$ (Poisson r.v. with
parameter $\lambda_en$), and the total number of legitimate
users in the network as $|\Lc|$ (Poisson r.v. with
parameter $\lambda n= n$). Let the total area in which the
eavesdroppers make it impossible to reach
a highway be $A_{\Ec}$. Clearly, $A_{\Ec}\leq A |\Ec|$.
Let us further denote the number of legitimate users in an
area of $A|\Ec|$ as $|\Lc_{A|\Ec|}|$. Then, using the Chebyshev
inequality (please refer to Lemma~\ref{thm:Lemma8:Chebyshev} in
Appendix~\ref{sec:Appendix1}), we obtain
\begin{eqnarray}\label{eq:Lemma5eq1}
|\Ec| &\leq& (1+\epsilon)\lambda_e n, \nonumber \\
|\Lc| &\geq& (1-\epsilon)n, \\
|\Lc_{A|\Ec|}| &\leq& (1+\epsilon)A |\Ec|, \nonumber
\end{eqnarray}
for any $\epsilon\in(0,1)$ with high probability
(as $n\to\infty$). We denote the fraction of users
that can not transmit to highways due to eavesdroppers
as $F$ which can be upper bounded by
\begin{equation}
F \leq \frac{|\Lc_{A|\Ec|}|}{|\Lc|} \leq
\frac{(1+\epsilon)^2 (2f_ed+1)^2c^2 \lambda_e n}
{(1-\epsilon) n} \to 0
\end{equation}
with high probability (as $n\to\infty$). The first inequality
follows since the eavesdroppers can have intersecting secrecy regions,
the second inequality follows from~\eqref{eq:Lemma5eq1}, and the
limit holds as $d = \kappa''\log(n)$ and
$ \lambda_e = o\left((\log n)^{-2}\right)$.
This argument shows that almost all of the nodes
are connected to the highways as $n\to\infty$.

Similar conclusion can be made for the final destination
nodes: Any given destination node can w.h.p. receive
data from the highways securely with a rate of
$\Omega\left((\log n)^{-3-\alpha}\right)$.
\end{IEEEproof}

Now we are ready to state our main result.

\begin{theorem}\label{thm:Theorem1}
If the legitimate nodes have unit intensity ($\lambda=1$) and the
eavesdroppers have an intensity of
$\lambda_e=o\left((\log n)^{-2}\right)$
in an extended network,
almost all of the nodes can achieve a secure rate of
$\Omega\left(\frac{1}{\sqrt{n}}\right)$ with high probability.
\end{theorem}

\begin{IEEEproof}
In our multi-hop routing scheme, each user has a
dedicated route (due to the time division
scheme described below) with each hop sending the message to the
next hop over $N$ channel uses.
The secrecy encoding at each
transmitter is designed assuming an eavesdropper on the
boundary of the secrecy zone and listening to this hop
(observations of length $N$) only. This way, a transmitter
can achieve the rate reported in Lemma~\ref{thm:Lemma1:SecureRate}.
Then, we can argue that this secrecy encoding
scheme will ensure secrecy from an
eavesdropper that listens to the transmissions of every hop
due to Lemma~\ref{thm:Lemma2:MultihopSecurity}.

Now, the main result follows by Lemma~\ref{thm:Lemma4:HighwayRate} and
Lemma~\ref{thm:Lemma5:AccessRate} by utilizing a time division approach.
That is the total transmission time of the network is divided
into four phases, as shown in Fig.~$1$.
During the first phase, the sources that are
not affected by eavesdroppers (i.e., almost all of them due to
Lemma~\ref{thm:Lemma5:AccessRate}) will w.h.p. transmit their messages to
the closest highway entry point. Then, the secret messages
of all nodes are carried through the horizontal
highways and then the vertical highways
(Lemma~\ref{thm:Lemma4:HighwayRate}). During the
final phase, the messages are delivered from the highways to almost
all of the destinations (Lemma~\ref{thm:Lemma5:AccessRate}).
Hence, by Lemma~\ref{thm:Lemma4:HighwayRate} and
Lemma~\ref{thm:Lemma5:AccessRate}, as the secrecy rate
scaling per node is limited by the transmissions
on the highway, we can see that almost all of
the nodes achieve a secure rate of
$\Omega\left(\frac{1}{\sqrt{n}}\right)$ with high probability.
This concludes the proof.
\end{IEEEproof}

Few remarks are now in order.

1) The expected number of legitimate nodes is
$n$, whereas the expected number of eavesdroppers is
$n_e=o(n(\log n)^{-2})$ in this extended network.
Note that $n_e$ satisfies $n_e=O(n^{1-\epsilon})$ for
any $\epsilon>0$, and hence network can endure eavesdroppers
as long as total number of eavesdroppers scale slightly
lower than that of legitimate nodes.

2) Utilizing the upper bound of~\cite{Gupta:The00} for the capacity
of wireless networks, we can see that Theorem~\ref{thm:Theorem1}
establishes the achievability of the same \textbf{optimal scaling law} with
and without security constraints. It is worth noting that, in our
model, the interference is considered as noise at the legitimate
receivers. As shown in~\cite{Ozgur:Hierarchical07}, more
sophisticated cooperation strategies achieve the same throughput for
the case of extended networks with $\alpha\geq 3$. This leads to the
conclusion that cooperation in the sense
of~\cite{Ozgur:Hierarchical07} does not increase the secrecy
capacity when $\alpha\geq 3$ and $\lambda_e=o\left((\log n)^{-2}\right)$.

3) \textbf{$\lambda_e=o(1)$ is tolerable if each node shares
key only with the closest highway member.}
If each node can share a secret key
with \emph{only} the closest highway member, then the
proposed scheme can be combined with a one-time pad scheme
(see, e.g.,~\cite{Vernam:OneTimePad26}
and~\cite{Shannon:Communication49})
for accessing the highways, which results in the same scaling
performance for any $\lambda_e\to 0$ as $n\to \infty$.

4) \textbf{Can network endure $\lambda_e=o(1)$ without key sharing?}
Note that in our percolation theory result, we have chosen
squares of side length $c$ (edge length in the square lattice
was $c\sqrt{2}$, see Fig.~$3$) satisfying $c\geq c_0$ to make
$pq$ sufficiently large in order to have $p'> \frac{5}{6}$
for Lemma~\ref{thm:Lemma9:Percolation}. We remark that
for independent percolation with edge probability $p'$ in a
random grid, for any $\gamma\in(0,1)$, $\exists p^*(\gamma)$ such
that for $p'>p^*(\gamma)$, the random grid contains a connected
component of at least $\gamma n^2$ vertices (see,
e.g.,~\cite[Theorem 3.2.2]{FranceschettiAndMeester:07}).
Thus, as long as $\lambda_e=o(1)$,
for some $\epsilon',\epsilon^*>0$, we can choose a very large,
but constant, $c$ (to make sure that $pq$ is very close to $1$)
to have $p'=1-\epsilon'>p^*(1-\epsilon^*)$, which implies that
there are w.h.p. $(1-\epsilon^*) n^2$ connected vertices. Therefore,
we conjecture that, for any given $\epsilon>0$ and for $\lambda_e=o(1)$,
per-node secure throughput of $\Omega(1/\sqrt{n})$ is achievable for
$(1-\epsilon)$ fraction of nodes (we conjecture that these are the nodes
that have constant distances to highways).

We now focus on the dense network scenario.
The stochastic node distribution for this scenario
can be modeled by assuming that the legitimate and eavesdropper
nodes are distributed as Poisson point processes of
intensities $\lambda=n$ and $\lambda_e$, respectively,
over a square region of unit area.
The proposed scheme in the previous section
can be utilized for this topology and the same scaling
result can be obtained for dense networks as formalized in
the following corollary.

\begin{corollary}\label{thm:Dense1}
Under the stochastic modeling of node distribution (Poisson
point processes) in a dense network (on a unit area region)
with the path loss model (with $\alpha >2$),
if the legitimate nodes have an intensity of $\lambda=n$
and the eavesdropper intensity satisfies
$\frac{\lambda_e}{\lambda}=o\left((\log n)^{-2}\right)$,
then almost all of the nodes can simultaneously achieve
a secure rate of
$\Omega\left(\frac{1}{\sqrt{n}}\right)$.
\end{corollary}
\begin{IEEEproof}
The claim can be proved by following the same steps of the
proof of Theorem~$\ref{thm:Theorem1}$ with scaling the transmit
power from $P$ to $\frac{P}{(\sqrt{n})^\alpha}$ at each transmitter,
and scaling each distance parameter by dividing with $\sqrt{n}$.
Note that, with these scalings, signal to interference and noise
ratio (SINR) calculations and percolation results remain unchanged.
\end{IEEEproof}


\section{The Ergodic Fading Model}
\label{sec:ErgodicFading}

We now focus on the ergodic fading model
described in Section~\ref{sec:SystemModel2} and utilize
the ergodic interference alignment for secrecy.
Frequency selective slow fading channels are studied
in~\cite{Koyluoglu:Interference}, where each symbol time $t=1,\cdots,N$
corresponds to $F$ frequency uses and the channel states of
each sub-channel remain constant for a block of
$N'$ channel uses and i.i.d. among $B$ blocks
($N=N'B$). For such a model, one can obtain the following high SNR
result by utilizing the interference alignment
scheme~\cite{Cadambe:Interference08}.

\begin{theorem}[Theorem 3 of~\cite{Koyluoglu:Interference}]\label{thm:Fading2}
For $n$ source-destination pairs with $n_e$ number of
external eavesdroppers, a secure DoF of
$\eta=\left[\frac{1}{2}-\frac{1}{n}\right]^+$ per
frequency-time slot is achievable at each user in the
ergodic setting, in the absence of the eavesdropper CSI,
for sufficiently high SNR, $N$, and $F$.
\end{theorem}
This interference alignment scheme is shown to achieve a secure
DoF of $\left[\frac{1}{2}-\frac{n_e}{n}\right]^+$ per orthogonal dimension at each
user when all the eavesdroppers collude~\cite{Koyluoglu:OntheEffect10}.
Remarkably, with this scheme, the network is secured against
colluding eavesdroppers and only a statistical knowledge
of the eavesdropper CSI is needed at the network users.
However, the proposed scheme only establishes a high SNR
result in terms of secure DoF per user. In addition,
the stated DoF gain is achieved in the limit of large number of
sub-channels, which is unrealistic in practice for large
number of users, $n$. (The result is achieved
when the design parameter $m$ gets large, where
$F=\Omega (m^{n^2})$~\cite{Koyluoglu:Interference},
\cite{Koyluoglu:OntheEffect10}.)

Providing secure transmission guarantees for users at
any SNR with finite number of dimensions is of definite interest.
In this section, we utilize the ergodic interference alignment
scheme~\cite{Nazer:Ergodic09} to satisfy this quality of service
(QoS) requirement at the expense of large coding delays.
Ergodic interference alignment can be summarized as
follows. Suppose that we can find some time indices
in $\{1,\cdots,N\}$, represented by
$t_1,t_2,\cdots$ and their complements
$\tilde{t}_1,\tilde{t}_2,\cdots$, such that
$h_{i,i}(t_m)=h_{i,i}(\tilde{t}_m)$, $\forall i\in\Kc$,
and $h_{i,j}(t_m)=-h_{i,j}(\tilde{t}_m)$,
$\forall i,j\in\Kc$ with $i\neq j$, for
$m=1,2,\cdots,N_1$. Now, consider that we sent the
same codeword over the resulting channels, i.e.,
we set $X_i(t_m)=X_i(\tilde{t}_m), \forall m$. Then,
by adding the observations seen by destination $i$
for these two time instance sequences, the effective
channel can be represented as
\begin{eqnarray}
\tilde{Y}_i (t_m) = 2 h_{i,i}(t_m) X_i(t_m)
+ Z_i(t_m) + Z_i(\tilde{t}_m),
\end{eqnarray}
whereas the eavesdropper $e$ observes
\begin{eqnarray}
\tilde{\Ym}_e (t_m) = \sum\limits_{i=1}^n
\left[
  \begin{array}{c}
    h_{i,e}(t_m) \\
    h_{i,e}(\tilde{t}_m) \\
  \end{array}
\right]
X_i(t_m)
+
\left[
  \begin{array}{c}
    Z_e(t_m) \\
    Z_e(\tilde{t}_m) \\
  \end{array}
\right],
\end{eqnarray}
for $m=1,2,\cdots,N_1$. Remarkably, while the interference
is canceled for the legitimate users, it still exists
for the eavesdropper, whose effective channel becomes
multiple access channel with single input multiple output
antennas (SIMO-MAC). By taking advantage of this phenomenon
together with exploiting the ergodicity of the channel,
secure transmission against each eavesdropper
is made possible at each user for any SNR
(depending on the underlying fading processes)
as reported in the following theorem, which is
the main result of this section.

\begin{theorem}\label{thm:Fading3}
For $t=1,2,\cdots$, let
\begin{eqnarray}\label{eq:Fading1}
\tilde{Y}_i (t) \triangleq 2 h_{i,i}(t) X_i(t)
+ Z_i(t) + \tilde{Z}_i(t),
\end{eqnarray}
\begin{eqnarray}\label{eq:Fading2}
\tilde{\Ym}_e (t) \triangleq \sum\limits_{i=1}^n
\tilde{\Hm}_{i,e} (t)
X_i(t)
+\tilde{\Zm}_{e} (t),
\end{eqnarray}
$\tilde{\Hm}_{i,e} (t) \triangleq
[h_{i,e}(t) \: \tilde{h}_{i,e}(t)]^T$, and
$\tilde{\Zm}_{e} (t) \triangleq
[Z_{e}(t) \: \tilde{Z}_{e}(t)]^T$,
where, $\forall i\in \Kc$ and  $\forall e\in \Ec$,
$\tilde{Z}_i$ and $\tilde{Z}_e$ are i.i.d. as
$Z_i$ and $Z_e$, respectively;
and $\tilde{h}_{i,e}$ is i.i.d. as $h_{i,e}$.
Then, source destination pair $i\in\Kc$
can achieve the secret rate
\begin{eqnarray}
R_i &=& \bigg[\frac{1}{2} E[I(X_i;\tilde{Y}_i|\Hm)]
-\frac{1}{2n} E[I(X_1,\cdots,X_n;\tilde{\Ym}_e|\Hm,\Hm_e)]\bigg]^+,
\end{eqnarray}
on the average, where the expectations are over
underlying fading processes.
\end{theorem}

\begin{IEEEproof}
We first need to quantize the channel gains to have a finite
set of possible matrices. (These steps are given
in~\cite{Nazer:Ergodic09} and provided here for completeness.)
Let $\epsilon'>0$.
Choose $\tau>0$ such that
$\Prob\{\cup_{i,j} \{ |h_{i,j}| > \tau \} \} \leq \epsilon'$.
This will ensure a finite quantization set.
For $\gamma>0$, the $\gamma$-quantization of $h_{i,j}$
is the point among $\gamma(\ZZ + j \ZZ)$ that is closest
to $h_{i,j}$ in Euclidean distance. The $\gamma$-quantization of
channel gain matrix $\Hm(t)$ is denoted by $\Hm_{\gamma}(t)$,
where each entry is $\gamma$-quantized. Thus, $\gamma$-quantized
channel alphabet $\Hc_{\gamma}$ has size satisfying
$(\frac{\sqrt{2}\tau}{\gamma})^{2n^2} \leq |\Hc_{\gamma}|
\leq (\frac{2\tau}{\gamma})^{2n^2}$.
We denote each channel type with $\Hm_{\gamma}^b$, for
$b=1,\cdots,B=|\Hc_{\gamma}|$. The complement of the
channel $\Hm_{\gamma}^b$ is denoted by
$\Hm_{\gamma}^{\tilde{b}}$, whose diagonal elements are the
same as $\Hm_{\gamma}^b$ and the remaining elements
are negatives of that of $\Hm_{\gamma}^b$.

We next utilize strong typicality~\cite{ThomasAndCover:91}
to determine the number of channel uses for each type.
Consider any i.i.d. sequence of quantized channel matrices
$\Hm_{\gamma}(1),\cdots,\Hm_{\gamma}(N)$.
Such a sequence is called $\delta$-typical, if
\begin{eqnarray}
N (\Prob\{\Hm_{\gamma}^b\} - \delta) \leq
\# \{\Hm_{\gamma}^b|\Hm_{\gamma}(1),\cdots,\Hm_{\gamma}(N)\}
\leq N (\Prob\{\Hm_{\gamma}^b\} + \delta ),
\end{eqnarray}
where $\#\{.|.\}$ operator gives the number of blocks of each type.
The set of such strong typical sequences is denoted by
$\Ac_{\delta}^{(N)}$. By the strong law of large numbers,
we choose sufficiently large $N$ to have
$\Prob\{\Ac_{\delta}^{(N)}\} \geq 1-\epsilon'$.

Assuming that the realized sequence of quantized channel
gain matrices, i.e.,
$\Hm_{\gamma}(1),\cdots,\Hm_{\gamma}(N)$,
is $\delta$-typical,
we use the first $N_b\triangleq N (\Prob\{\Hm_{\gamma}^b\} - \delta )$
channel uses for each channel type $b$.
This causes a loss of at most $2\delta NB$ channel uses
out of $N$,
which translates to a negligible rate loss.
With again a negligible loss in the rate, we
choose each $N_b$ as even. Note that the complement
block of $b$ is $\tilde{b}$, which lasts for
$N_{\tilde{b}}=N_b$ channel uses, as
$\Prob\{\Hm_{\gamma}^b\}=\Prob\{\Hm_{\gamma}^{\tilde{b}}\}$.

We now describe the coding scheme, which can be
viewed as an ergodic interference alignment coding
scheme with a secrecy pre-coding.
For each secrecy codebook in the ensemble
of transmitter $i$, we generate $2^{N(R_i+R_i^x)}$
sequences each of length $\sum\limits_{b=1}^B \frac{N_b}{2}$,
where entries are chosen such that they
satisfy the long term average power constraint of $P$.
We assign each codeword to $2^{NR_i}$ bins each with
$2^{NR_i^x}$ codewords. Given $w_i$, transmitter randomly chooses
a codeword in bin $i$ according to the uniform distribution,
which is denoted by $\Xm_i(w_i,w_i^x)$, where $w_i^x$
is the randomization index to confuse the eavesdroppers.
The codeword is then divided into $B$ blocks each with
a length of $\frac{N_b}{2}$ symbols.
The codeword of block $b$ is denoted by
$\{X_i^b(t), t=1,\cdots,\frac{N_b}{2}\}$ and is
repeated during the last $\frac{N_b}{2}$ channel
uses of the block $\tilde{b}$, i.e.,
$X_i^b(t)=X_i^{\tilde{b}}(\frac{N_b}{2}+t)$,
for $t=1,\cdots,\frac{N_b}{2}$.
The channel gains, additive noises, and the received
symbols is denoted with the same block, i.e., channel
type, notation. Here, the effective channels during
block $b$ is given by
\begin{eqnarray}\label{eq:Fading3}
\tilde{Y}_i^b (t) = 2 h_{i,i}^b(t) X_i^b(t)
+ Z_i^b(t) + Z_i^{\tilde{b}}\left(\frac{N_b}{2}+t\right),
\end{eqnarray}
and
\begin{eqnarray}\label{eq:Fading4}
\tilde{\Ym}_e^b (t) = \sum\limits_{i=1}^n
\left[
  \begin{array}{c}
    h_{i,e}^b(t) \\
    h_{i,e}^{\tilde{b}}\left(\frac{N_b}{2}+t\right) \\
  \end{array}
\right]
X_i^b(t)
+
\left[
  \begin{array}{c}
    Z_e^b(t) \\
    Z_e^{\tilde{b}}\left(\frac{N_b}{2}+t\right) \\
  \end{array}
\right],
\end{eqnarray}
for $t=1,2,\cdots,\frac{N_b}{2}$.

We essentially code over the above two fading channels
seen by destinations and eavesdroppers. Here, to satisfy
both the secrecy and the reliability constraints, we
choose the rates as follows.
\begin{eqnarray}
R_i &=& \frac{1}{2} E[I(X_i;\tilde{Y}_i|\Hm)]
-\frac{1}{2n} E[I(X_1,\cdots,X_n;\tilde{\Ym}_e|\Hm,\Hm_e)] - \epsilon  \\
R_i^x &=& \frac{1}{2n} E[I(X_1,\cdots,X_n;\tilde{\Ym}_e|\Hm,\Hm_e)],
\end{eqnarray}
where the expectation is over the ergodic channel fading, and
the channel outputs $\tilde{Y}_i$ and $\tilde{\Ym}_e$
are given by the transformations
\eqref{eq:Fading1} and \eqref{eq:Fading2}, respectively.

For any $\epsilon>0$, we choose sufficiently small $\delta$.
Then, in the limit of $N\to \infty$, $\tau\to \infty$,
$\gamma\to 0$, each legitimate receiver $i$ can decode
$W_i$ and $W_i^x$ with high probability (covering a-typical
behavior of the channel sequence as well) as
\begin{eqnarray}
R_i + R_i^x = \frac{1}{2} E[I(X_i;\tilde{Y}_i|\Hm)] - \epsilon,
\end{eqnarray}
where $\epsilon$ covers for quantization errors and unused
portion of the channel uses.

For the secrecy constraint we first consider each
expression on the right hand side of the following equality.
\begin{eqnarray}\label{eq:Fading5}
\frac{1}{N} I(W_{\Kc};\Ym_e,\Hm,\Hm_e)
&=& \frac{1}{N} I(W_{\Kc},W_{\Kc}^x;\Ym_e,\Hm,\Hm_e)
+ \frac{1}{N} H(W_{\Kc}^x|W_{\Kc},\Ym_e,\Hm,\Hm_e) \nonumber\\
&& {-}\: \frac{1}{N} H(W_{\Kc}^x|W_{\Kc},\Hm,\Hm_e),
\end{eqnarray}
where we denote $W_{\Kc}\triangleq \{W_i, \forall i\in\Kc\}$
and $W_{\Kc}^x\triangleq \{W_i^x, \forall i\in\Kc\}$.

We have
$$\frac{1}{N} I(W_{\Kc},W_{\Kc}^x;\Ym_e,\Hm,\Hm_e)
= \frac{1}{N} I(W_{\Kc},W_{\Kc}^x;\Ym_e|\Hm,\Hm_e)$$
\begin{eqnarray}\label{eq:Fading6}
&\stackrel{(a)}{\leq}& \frac{1}{N}
I(\{X_i^b(t), \forall i,b,t\};\{\tilde{\Ym}_e^b(t), \forall b,t\}|\Hm,\Hm_e) \nonumber \\
&\stackrel{(b)}{\leq}&
\frac{\sum\limits_{b=1}^B\frac{N_b}{2}}{N}
\left( E[I(X_1,\cdots,X_n;\tilde{\Ym}_e|\Hm,\Hm_e)] - \epsilon_1 \right) \nonumber \\
&\stackrel{(c)}{=}&
\frac{(1-\epsilon_2)}{2}
\left( E[I(X_1,\cdots,X_n;\tilde{\Ym}_e|\Hm,\Hm_e)] - \epsilon_1 \right) \nonumber \\
&\leq&
\frac{1}{2} E[I(X_1,\cdots,X_n;\tilde{\Ym}_e|\Hm,\Hm_e)] + \epsilon_1 \epsilon_2,
\end{eqnarray}
where (a) is due to the coding scheme and the data
processing inequality, (b) is due to ergodicity with some
$\epsilon_1\to 0$ as $N\to \infty$, (c) is due to
unused portion of channel uses with some
$\epsilon_2\to 0$ as $N\to \infty$.

Secondly, due to the ergodicity and the symmetry among tranmitters,
the rate assignment implies the following:
The rates satisfy
\begin{eqnarray}
\sum\limits_{i\in\Sc} R_i^x \leq \frac{1}{2}
E[I(X_{\Sc};\tilde{\Ym}_e|X_{\Kc-\Sc},\Hm,\Hm_e)],
\end{eqnarray}
for any $\Sc\subseteq\Kc$. (Please refer to
Lemma 8 of~\cite{Koyluoglu:Interference} for details.)
Thus, the randomization indices $W_{\Kc}^x$ can be decoded
at the eavesdropper $e$ given the bin indices $W_{\Kc}$.
Then, utilizing Fano's inequality and averaging over the
ensemble of the codebooks, we have
\begin{eqnarray}\label{eq:Fading7}
\frac{1}{N} H(W_{\Kc}^x|W_{\Kc},\Ym_e,\Hm,\Hm_e) \leq \epsilon_3,
\end{eqnarray}
with some $\epsilon_3\to 0$ as $N\to \infty$.

Third, as $W_{\Kc}^x$ is independent of $\{W_{\Kc},\Hm,\Hm_e\}$
and as each $W_i^x$ is independent, we have
\begin{eqnarray}\label{eq:Fading8}
\frac{1}{N} H(W_{\Kc}^x|W_{\Kc},\Hm,\Hm_e)
= \frac{1}{N} H(W_{\Kc}^x)
= \frac{1}{N} \sum\limits_{i=1}^n H(W_i^x)
= \frac{1}{N} \sum\limits_{i=1}^n NR_i^x
= \frac{1}{2} E[I(X_1,\cdots,X_n;\tilde{\Ym}_e|\Hm,\Hm_e)].
\end{eqnarray}

Finally, using \eqref{eq:Fading6}, \eqref{eq:Fading7}, and
\eqref{eq:Fading8} in \eqref{eq:Fading5}, we obtain
\begin{eqnarray}
\frac{1}{N} I(W_{\Kc};\Ym_e,\Hm,\Hm_e) \leq \epsilon,
\end{eqnarray}
which implies that
\begin{eqnarray}
\frac{1}{N} I(W_i;\Ym_e,\Hm,\Hm_e) \leq \epsilon, \forall i\in\Kc
\end{eqnarray}
with some $\epsilon\to 0$ as $N\to \infty$,
which establishes the claim.
\end{IEEEproof}

Note that for i.i.d. complex Gaussian input distribution, i.e.,
when $X_i(t)~\sim\Cc\Nc(0,P), \forall i,t$, the proposed
scheme achieves
\begin{eqnarray}\label{eq:AchievedRate}
R_i &=& \Bigg[\frac{1}{2}E\left[\log\left(1+ \frac{2P|h_{i,i}|^2}{N_0}\right)\right]
- \frac{1}{2n} E\left[\log \det \left(\Id_2 +
\frac{P}{N_0} \sum\limits_{i=1}^n \tilde{\Hm}_{i,e} \tilde{\Hm}_{i,e}^*
\right)\right]
\Bigg]^+,
\end{eqnarray}
for user $i\in\Kc$.
Here, for any non-degenerate fading distribution, e.g.,
Rayleigh fading where
$h_{i,k}~\sim \Cc\Nc(0,1), \forall i\in\Kc, \forall k\in\Kc \cup \Ec$,
the second term of~\eqref{eq:AchievedRate} diminishes as $n$
gets large. In particular, as $n\to\infty$,
$R_i$ scales as
\begin{eqnarray}
R_i &=& \Bigg[\frac{1}{2}E\left[\log\left(1+ \frac{2P|h_{i,i}|^2}{N_0}\right)\right]
- \frac{O(\log (n))}{n} \Bigg]^+,\nonumber
\end{eqnarray}
and hence we can say that each user can achieve at least
a positive constant secure rate for any given SNR for
sufficiently large $n$.
(Please refer to Appendix~\ref{sec:Appendix2}.)

To quantify the behavior of the scheme in the high
SNR regime, we now focus on the achievable secure DoF per user,
which can be characterized by dimension counting arguments.
The proposed scheme achieves $\eta=\left[\frac{1}{2}-\frac{1}{n}\right]^+$
secure DoF per user for any given non-degenerate fading model.
(This can be shown by dividing both sides of~\eqref{eq:AchievedRate}
with $\log \textrm{SNR}$ and taking the limit
$\textrm{SNR}\to\infty$ for any given $n$.)
Note that the pre-log gain of the proposed scheme is the
same as that of~\cite{Koyluoglu:Interference}.
But, remarkably, ergodic interference alignment allows us to
attain secrecy at any SNR by only requiring a statistical
knowledge of the eavesdropper CSI. We note that this gain
is obtained at the expense of large coding delay (at least
exponential in the number of users).


\section{Eavesdropper Collusion}
\label{sec:Colluding}

In a more powerful attack, eavesdroppers can \emph{collude}, i.e., they
can share their observations. Securing information in such a scenario will
be an even more challenging task compared to non-colluding
case~\cite{Pinto:Wireless09,Goel:Secret05}.
Interestingly, even with colluding eavesdroppers, we
show that the scaling result for the path loss model
remains the same with the proposed multi-hop scheme with
almost the same eavesdropper intensity requirement.

\begin{theorem}\label{thm:ColludingPathLoss}
If the legitimate nodes have unit intensity ($\lambda=1$) and the
colluding eavesdroppers have an intensity of $\lambda_e=O\left((\log
n)^{-2-\rho}\right)$ for any $\rho>0$ in an extended network, almost
all of the nodes can achieve a secure rate of
$\Omega\left(\frac{1}{\sqrt{n}}\right)$ under the static
path loss channel model.
\end{theorem}

\begin{IEEEproof}
Please refer to Appendix~\ref{sec:Appendix3}.
\end{IEEEproof}

We note that, in the colluding eavesdropper scenario, the result
requires only a slightly modified eavesdropper intensity condition
compared to the non-colluding case.
Also, for the highway construction of the non-colluding case,
the secrecy zone with an area of $(2df_{e}+1)^2c^2$ with
$f_e>\sqrt{2}$ was sufficient. However, for the
colluding eavesdropper scenario, legitimate nodes need to know whether
there is an eavesdropper or not within the first layer zone, which has an
area of $(2df_{l_1}+1)^2c^2$ with $f_{l_1}=\delta' \log(n)$, where
$\delta'$ can be chosen arbitrarily small
(see~\eqref{eq:Appendix3eq2}). Hence, securing the network against
colluding eavesdroppers requires more information regarding
the eavesdroppers compared to the non-colluding case.
But, remarkably, the optimal scaling law (see~\cite{Franceschetti:Closing07})
is achieved even when the eavesdroppers collude under these
assumptions.

For the ergodic fading model, the eavesdropper collusion decreases
the achievable performance. Let us
add independent observations to the received vector given
in~\eqref{eq:Fading2} of Theorem~\ref{thm:Fading3}
according to eavesdropper collusion and
denote colluding eavesdroppers' observations by $\tilde{\Ym}_{e^*}$
for $e^*\in\Ec^*\triangleq\{e_1^*,e_2^*,\cdots\}$.
For example, if $e_1$ and $e_2$ colludes, their
cumulative observations is denoted by $\tilde{\Ym}_{e_1^*}$
(SIMO-MAC with $4$ receive antennas).
In such a scenario, the proposed scheme can be used to
achieve the following rate.

\begin{corollary}
For a given eavesdropper collusion set $\Ec^*$,
source-destination pair $i\in\Kc$ achieves
the following rate with the proposed ergodic
interference alignment scheme for the ergodic
fading channel model:
\begin{eqnarray}\label{eq:AchievedRateColluding}
R_i = \min\limits_{e^*\in\Ec^*}
\bigg[\frac{1}{2} E[I(X_i;\tilde{Y}_i|\Hm)]
- \frac{1}{2n} E[I(X_1,\cdots,X_n;\tilde{\Ym}_{e^*}|\Hm,\Hm_e)] \bigg].
\end{eqnarray}
\end{corollary}

Note that the proposed scheme achieves
$\eta=\left[\frac{1}{2}-\frac{n_e}{n}\right]^+$ secure
DoFs per user for non-degenerate fading distributions
when all the eavesdroppers collude.
(This can be shown from~\eqref{eq:AchievedRateColluding} by setting
$\Ec^*=\Ec$, choosing the input distribution as i.i.d.~$\Cc\Nc(0,P)$,
dividing both sides by $\log \textrm{SNR}$, and taking the limit
$\textrm{SNR}\to\infty$ for any given $n_e$ and $n$.)


\section{Conclusion}
\label{sec:Conclusion}

In this work, we studied the scaling behavior of the capacity of
wireless networks under secrecy constraints.
For extended networks with the path loss model (the exponent
is assumed to satisfy $\alpha>2$), the legitimate nodes and
eavesdroppers were assumed to be randomly placed in the
network according to Poisson point processes of intensity
$\lambda=1$ and $\lambda_e$, respectively. It is shown that, when
$\lambda_e=o\left((\log n)^{-2}\right)$, almost all of the nodes
achieve a secure rate of $\Omega\left(\frac{1}{\sqrt{n}}\right)$,
showing that securing the transmissions does not entail a loss in the
per-node throughput for our model, where transmissions from other users
are considered as noise at receivers.
Our achievability argument is based on novel secure multi-hop
forwarding strategy where forwarding nodes are chosen such that no
eavesdroppers exist in appropriately constructed {\em secrecy zones}
around them and independent randomization is employed in each hop.
Tools from percolation theory were used to establish the existence
of a sufficient number of {\em secure highways} allowing for network
connectivity. Finally, a time division approach was used to
accomplish an end-to-end secure connection between almost all
source-destination pairs.
The same scaling result is also obtained for the dense network
scenario when
$\frac{\lambda_e}{\lambda}=o\left((\log n)^{-2}\right)$.
We note that, in the proposed scheme, we assumed that nodes know whether
an eavesdropper exist in a certain zone (secrecy zone) or not.
An analysis of a more practical scenario, in which legitimate nodes have no
(or more limited) eavesdropper location information, would be interesting.

We next focused on the ergodic fading model and
employed ergodic interference alignment scheme with an appropriate
secrecy pre-coding at each user. This scheme is shown to be capable of securing
each user at any SNR (depending on the underlying fading distributions),
and hence provides performance guarantees even for the
finite SNR regime compared to previous work. For the high SNR scenario,
the scheme achieves $[\frac{1}{2}-\frac{1}{n}]^+$ secure DoFs
per orthogonal dimension at each user.
Remarkably, the results for the ergodic fading
scenarios do not require eavesdropper CSI at the legitimate users,
only a statistical knowledge is sufficient. However, this gain is
obtained at the expense of large coding delays.

Lastly, the effect of the eavesdropper collusion is analyzed.
It is shown that, for the path loss model, the same per-node throughput
scaling, i.e., $\Omega\left(\frac{1}{\sqrt{n}}\right)$, is achievable under
almost the same eavesdropper intensity requirement. For the fading model,
the proposed model is shown to endure various eavesdropper collusion
scenarios. In particular, when all the eavesdroppers collude, a
secure DoF of $[\frac{1}{2}-\frac{n_e}{n}]^+$ is shown to be
achievable.

We list several future directions here: 1) Characterizing the full
trade-off between secure throughput vs. eavesdropper node intensity
is of definite interest.
2) We have not exploited cooperation techniques to enhance security in
this work. Cooperation in the sense of~\cite{Ozgur:Hierarchical07}
may be helpful. For example, in the extended network scenario,
hierarchical cooperation might increase the per-node throughput
for $\alpha<3$ or achieve the optimal throughput for
$\alpha \geq 3$ even with higher eavesdropper intensities.
In addition, cooperation for secrecy strategies
(see, e.g.,~\cite{Koyluoglu:Onthe08,Koyluoglu:Cooperative}
and references therein) may be beneficial in enhancing
the scaling results.
3) A uniform rate per user is considered in this work.
Arbitrary traffic pattern can be considered for users with
distinct quality of service constraints.
4) Eavesdroppers are assumed to be passive
(they only listen the transmissions). An advanced attack
might include active eavesdroppers, which may jam
the wireless channel. Securing information in such
scenarios is an interesting avenue for further research.


\appendices


\section{Lemmas used in Section~\ref{sec:PathLoss}}
\label{sec:Appendix1}

\begin{lemma}[Theorem 7.65,~\cite{Grimmett:99}]\label{thm:Lemma6:DependentPercolation}
Let $d,k\geq1$. Consider random variables $Y_x$ and $Z^\pi_x$
taking values in $\{0,1\}$, for $x\in\ZZ^d$.
Denote $Z^\pi=\{Z^\pi_x:x\in \ZZ^d\}$
as a family of independent random variables satisfying
$\Prob\{Z^\pi_x=1\}=1-\Prob\{Z^\pi_x=0\}=\pi$.
Also, denote Euclidean distance in $\ZZ^d$ as $d(\cdot,\cdot)$.

If $Y=\{Y_x:x\in \ZZ^d\}$ is a $k$-dependent family
of random variables, i.e., if any two sub-families
$\{Y_x:x\in\Ac\}$ and $\{Y_x':x'\in\Ac'\}$
are independent whenever $d(x,x')>k$,
$\forall x\in\Ac, \forall x'\in\Ac'$, such that
$$\Prob\{Y_x=1\}\geq \delta,\: \forall x\in\ZZ^d,$$
then there exist a family of independent random variables
$Z^{\pi(\delta)}$ such that $Y$ \emph{statistically dominates}
$Z^{\pi(\delta)}$, where ${\pi(\delta)}$ is a non-decreasing function
$\pi:[0,1]\to[0,1]$ satisfying
$\pi(\delta)\to 1$ as $\delta\to 1$.
\end{lemma}

\begin{IEEEproof}
The proof is given in~\cite{Liggett:Domination97},
where the authors also provide a construction of the
independent model. See also~\cite{Grimmett:99}.
\end{IEEEproof}

\begin{lemma}[Theorem 5,~\cite{Franceschetti:Closing07}]\label{thm:Lemma9:Percolation}
Consider discrete edge percolation with edge existence probability
$p$ on a square grid of size $m \times m$ (number of edges).
For any given $\kappa>0$, partition the area into
$\frac{m}{(\kappa \log m - \epsilon_m)}$ rectangles of
size $m \times (\kappa \log m - \epsilon_m)$, where $\epsilon_m=o(1)$
as $m\to\infty$ and is chosen to have integer number of rectangles.
Denote the maximal number of edge-disjoint left to right crossings of
the $i$th rectangle as $C_m^i$ and let $N_m\triangleq \min_i C_m^i$.
Then, $\forall \kappa>0$ and $\forall p\in(\frac{5}{6},1)$ satisfying
$\kappa \log (6 (1-p)) < -2$, $\exists \delta>0$ such that
\begin{eqnarray}
\lim\limits_{m\to\infty} \Prob\{ N_m \leq \delta \log m\} = 0.
\end{eqnarray}
\end{lemma}
\begin{IEEEproof}
The proof is given in~\cite[Appendix I]{Franceschetti:Closing07}.
See also~\cite[Theorem 4.3.9]{FranceschettiAndMeester:07}.
\end{IEEEproof}

\begin{lemma}\label{thm:Lemma7:Chernoff}
Consider a Poisson random variable $X$ of parameter
$\lambda$. Then,
\begin{equation}
P(X\geq x) \leq \frac{e^{-\lambda}(e\lambda)^x}{x^x},
\textrm{ for } x>\lambda.
\end{equation}
\end{lemma}

\begin{IEEEproof}
The proof follows by an application of the Chernoff bound.
Please refer to~\cite[Appendix II]{Franceschetti:Closing07}
or~\cite[Appendix]{FranceschettiAndMeester:07}.
\end{IEEEproof}

\begin{lemma}\label{thm:Lemma8:Chebyshev}
Consider a Poisson random variable $X$ of parameter
$\lambda$. Then, for any $\epsilon\in(0,1)$,
\begin{equation}
\lim_{\lambda \to\infty} P(X \leq (1-\epsilon)\lambda) = 0,
\end{equation}
and
\begin{equation}
\lim_{\lambda \to\infty} P(X \leq (1+\epsilon)\lambda) = 1.
\end{equation}
\end{lemma}

\begin{IEEEproof}
The proof follows by utilizing the Chebyshev's inequality.
\end{IEEEproof}


\section{$R_i>R$ for some constant $R>0$ in~\eqref{eq:AchievedRate} as $n\to\infty$}
\label{sec:Appendix2}

Consider that the statistics of $h_{i,e}$s are given by
1) $q\triangleq E[\Re\{h_{i,e}\}]+jE[\Im\{h_{i,e}\}]$ is a complex
number with finite real and imaginary parts, and
2) $s\triangleq E[|h_{i,e}|^2]$ is a finite real number,
$\forall i\in\Kc, e\in\Ec$.
Let us further assume that
$\Id_2 + \frac{P}{N_0} \sum\limits_{i=1}^n \tilde{\Hm}_{i,e} \tilde{\Hm}_{i,e}^*$
is a positive definite matrix. Focusing on the second term
of~\eqref{eq:AchievedRate}, we obtain
\begin{eqnarray}
\frac{1}{2n} E\left[\log \det \left(\Id_2 +
\frac{P}{N_0} \sum\limits_{i=1}^n \tilde{\Hm}_{i,e} \tilde{\Hm}_{i,e}^*
\right)\right]&\stackrel{(a)}{\leq}& \frac{1}{2n} \log \det \left(\Id_2 +
\frac{P}{N_0} \sum\limits_{i=1}^n E\left[\tilde{\Hm}_{i,e} \tilde{\Hm}_{i,e}^*\right]
\right) \\
&\stackrel{(b)}{=}& \frac{1}{2n} \log \left(
1+\frac{P}{N_0}2ns+\frac{P^2}{N_0^2}n^2(s^2-|q|^4)
\right) \\
&=& \frac{O(\log (n))}{n},
\end{eqnarray}
where (a) is due to Jensen's inequality as $\log \det (\cdot)$ function is
concave in positive definite matrices, and (b) follows
from
$$\tilde{\Hm}_{i,e} \tilde{\Hm}_{i,e}^*=
\left(
  \begin{array}{cc}
    |h_{i,e}|^2 & h_{i,e}\tilde{h}_{i,e}^* \\
    \tilde{h}_{i,e}h_{i,e}^* & |\tilde{h}_{i,e}|^2 \\
  \end{array}
\right),
$$
which implies
$$E\left[\tilde{\Hm}_{i,e} \tilde{\Hm}_{i,e}^*\right]=
\left(
  \begin{array}{cc}
    s & |q|^2 \\
    |q|^2 & s \\
  \end{array}
\right).
$$

Thus, the second term of~\eqref{eq:AchievedRate} becomes
insignificant, $o(1)$ as $n\to\infty$; and $\exists R>0$ such that
$R_i>R, \forall i\in\Kc$ for sufficiently large $n$.
Note that the assumption that
$\Id_2 + \frac{P}{N_0} \sum\limits_{i=1}^n \tilde{\Hm}_{i,e} \tilde{\Hm}_{i,e}^*$
is a positive definite matrix holds in the limit of large $n$ almost surely.
(Here, due to strong law of large numbers, the sum converges to
$nE\left[\tilde{\Hm}_{i,e} \tilde{\Hm}_{i,e}^*\right]$
with probability $1$.)


\section{Proof of Theorem~\ref{thm:ColludingPathLoss}}
\label{sec:Appendix3}

The proof follows along the same lines of the proof of Theorem~\ref{thm:Theorem1}
by generalizing the secrecy zone approach to multi-level zones,
where the area of each zone is carefully chosen to obtain a (statistically)
working bound for the SNR of the colluding eavesdropper.

In Fig.~$5$, we show the \emph{zones} around a transmitting square:
Zone of level $k$ for $k\in\{1,\cdots,L\}$
has an area of $A_{l_k}$, and the associated distance is denoted with
$f_{l_k} d c$ with some $f_{l_k}\geq 1$ and $f_{l_k}\geq f_{l_{k-1}}$.
Note that, we take $f_{l_k}$ as a design parameter.
We will choose $f_{l_k}$ differently, depending on whether a node is
forwarding data over a highway or accessing to/accessed by a highway.
Furthermore, $d$ and $f_{l_k}$ may depend on $n$, i.e.,
expected number of users.

We now provide generalization of Lemma 1 to the colluding eavesdropper case.
\begin{lemma}[Secure Rate per Hop]\label{thm:Lemma1:ColludingSecureRate}
In a communication scenario depicted in Fig.~$5$ (no eavesdroppers
in the first zone), the rate
\begin{equation}
R_{TR} = \frac{1}{(f_td)^2} \left[\log(1 +
\underline{\textrm{SNR}_{TR}}) - \log(1 +
\overline{\textrm{SNR}_{\Ec^*}})\right]^+,
\end{equation}
where
\begin{eqnarray}
\underline{\textrm{SNR}_{TR}}
&\triangleq & \frac{P (d+1)^{-\alpha}c^{-\alpha} (\sqrt{2})^{-\alpha}}
{N_o + P 8 (f_t)^{-\alpha}d^{-\alpha}c^{-\alpha} S(\alpha)},\\
S(\alpha) &\triangleq & \sum\limits_{i=1}^{\infty} i(i-0.5)^{-\alpha},\\
\overline{\textrm{SNR}_{\Ec^*}} &\triangleq & \frac{P(1+\epsilon)9c^{2-\alpha}
d^{-\alpha}}{N_0} \lambda_e d^2 \sum\limits_{k=2}^L
(f_{l_k})^2 (f_{l_{k-1}})^{-\alpha},\\
f_t &\geq & \frac{2(d+1)}{d},
\end{eqnarray}
is w.h.p. securely and simultaneously achievable between any
active transmitter-receiver pair if $f_{l_k}$ is chosen such that
\begin{equation}
\label{eq:Appendix3eq1} \lambda_e d^2 (f_{l_k})^2 \to \infty, \textrm{ as }
n\to\infty, \textrm{ for } k=2,3,\cdots.
\end{equation}
\end{lemma}
\begin{IEEEproof}
The steps of the proof are similar to that of Lemma~\ref{thm:Lemma1:SecureRate}.
Here, we need to derive a working upper bound for the colluding eavesdropper SNR.
In our case, secrecy is guaranteed assuming that the eavesdroppers are
located on the boundary of each level of zones. We first bound the number of
eavesdroppers at each level. We have
\begin{eqnarray}
A_{l_k} \leq (2df_{l_k}+1)^2 c^2 \leq 9d^2 (f_{l_k})^2 c^2,
\end{eqnarray}
as $d\geq1$ and $f_{l_k}\geq1$.
Hence, the number of eavesdroppers in layer $l_k$
can be bounded, using the Chebyshev's inequality
(see Lemma~\ref{thm:Lemma8:Chebyshev}), by
\begin{eqnarray}
|\Ec_{l_k}^*| \leq (1+\epsilon)\lambda_e 9 c^2 d^2 (f_{l_k})^2
\end{eqnarray}
w.h.p., for a given $\epsilon >0$, as long as we choose $f_{l_k}$ to satisfy
$$\lambda_e d^2 (f_{l_k})^2 \to \infty, \textrm{ as } n\to\infty.$$

Now, we place $|\Ec_{l_k}^*|$ number of eavesdroppers from
layer $k$ at distance $f_{l_{k-1}}dc$ for $k=2,3,\cdots$.
This is referred to as configuration $\Ec^*$. These colluding
eavesdroppers can do maximal ratio combining (this gives the best
possible SNR for them) to achieve the following SNR.

\begin{eqnarray}
\textrm{SNR}_{\Ec^*} &=& \frac{P \sum\limits_{k=2}^L |\Ec_{l_k}^*|
(f_{l_{k-1}})^{-\alpha} c^{-\alpha} d^{-\alpha}}{N_0} \nonumber\\
&\leq& \frac{P(1+\epsilon)9 c^{2-\alpha} d^{-\alpha}}{N_0}
\lambda_e d^2 \sum\limits_{k=2}^L (f_{l_k})^2 (f_{l_{k-1}})^{-\alpha}\nonumber\\
&\triangleq& \overline{\textrm{SNR}_{\Ec^*}}.
\end{eqnarray}

\end{IEEEproof}

Note that the challenge here is to choose $f_{l_k}$ such that
$\overline{\textrm{SNR}_{\Ec^*}}<\infty$, and at the same time
to satisfy (\ref{eq:Appendix3eq1}).
With some appropriate choices of these parameters, we
generalize Lemma~\ref{thm:Lemma4:HighwayRate} and
Lemma~\ref{thm:Lemma5:AccessRate} to the colluding eavesdropper
case.

\begin{lemma}[Rate per Node on the Highways]\label{thm:Lemma4:ColludingHighwayRate}
If $\lambda_e=O((\log n)^{-2})$,
each node on the constructed highways can transmit to
their next hop at a constant secure rate. Furthermore,
if the number of nodes each highway serves is $O(\sqrt{n})$,
each highway can w.h.p. carry a per-node
throughput of $\Omega\left(\frac{1}{\sqrt{n}}\right)$.
\end{lemma}
\begin{IEEEproof}

We show the result for
$\lambda_e=\Theta\left((\log n)^{-2}\right)$,
which will imply the desired result (as lowering
the eavesdropper density can not degrade the performance).
Consequently, there exists constants $\underline{\Lambda}$,
$\overline{\Lambda}$, and $n_1$ such that
\begin{equation}
\underline{\Lambda} (\log n)^{-2} \leq \lambda_e
\leq \overline{\Lambda} (\log n)^{-2}, \textrm{ for }
n\geq n_1,
\end{equation}
where $\underline{\Lambda} < \overline{\Lambda}$.

We choose each level of zones over the highways by setting
\begin{equation}\label{eq:Appendix3eq2}
f_{l_k} = \left(\frac{\delta}{9 \overline{\Lambda} c^2 d^2}\right)^{\frac{1}{2}}
(\log n)^{(\frac{\alpha}{2})^{k-1}}.
\end{equation}

Here,
\begin{eqnarray}
\lambda_e(2f_{l_1} d + 1)^2 c^2 &\leq&  \lambda_e 9(f_{l_1})^2 d^2 c^2 \\
&=& \lambda_e \frac{\delta (\log n)^2}{\overline{\Lambda}}\\
&\leq& \delta, \textrm{ for } n\geq n_1.
\end{eqnarray}
Therefore, due to our percolation result, i.e., Lemma~$3$, each
member of a given highway does not have any eavesdropper within
their first level secrecy zone as $\delta$ can be chosen
arbitrarily small. Now, as the above choice also satisfies
$$\lambda_e d^2 (f_{l_k})^2 \to \infty, \textrm{ as } n\to\infty, \textrm{ for } k=2,3,\cdots,$$
we can utilize Lemma~$1$ to achieve a secrecy rate of
\begin{equation}
R_{TR} = \frac{1}{(f_td)^2} \left[\frac{1}{2}\log(1 +
\underline{\textrm{SNR}_{TR}}) - \frac{1}{2}\log(1 +
\overline{\textrm{SNR}_{\Ec^*}})\right] .
\end{equation}

Now, we provide an upper bound for $\overline{\textrm{SNR}_{\Ec^*}}$.
First, note that our setup results in
$$(f_{l_k})^2 (f_{l_{k-1}})^{-\alpha} =
\left(\frac{\delta}{9 \overline{\Lambda} c^2
d^2}\right)^{\frac{2-\alpha}{2}}.$$
Hence,

\begin{eqnarray}
\overline{\textrm{SNR}_{\Ec^*}}
&=& \frac{P(1+\epsilon)9}{N_0}
\lambda_e  (L-1) \left(\frac{\delta}{9 \overline{\Lambda} }\right)^{\frac{2-\alpha}{2}}\\
&\leq& \frac{P(1+\epsilon)9}{N_0}
\overline{\Lambda} (\log n)^{-2}  (L-1)
\left(\frac{\delta}{9 \overline{\Lambda} }\right)^{\frac{2-\alpha}{2}},\nonumber\\
&&\: \: \textrm{ for } n\geq n_1 \\
&\to& 0, \textrm{ as } n\to \infty,
\end{eqnarray}
where the last step is due to the observation that
the number of levels can be upper bounded by
\begin{equation}
L-1\leq \frac{\log(\log n)}{\log(\frac{\alpha}{2})}.
\end{equation}

Therefore, there exists $n_2$ such that
for all $n\geq n_2$, the rate expression
satisfies $R_{TR}\geq R$ for some constant $R$.
The second claim follows from Lemma~$3$.

\end{IEEEproof}

\begin{lemma}[Access Rate to Highways]\label{thm:Lemma5:ColludingAccessRate}
Almost all source (destination) nodes can w.h.p. simultaneously
transmit (receive) their messages to (from) highways with a secure
rate of $\Omega\left((\log n)^{-3-\alpha}\right)$, if
$\lambda_e=O\left((\log n)^{-(2+\rho)}\right)$ for any $\rho>0$.
\end{lemma}

\begin{IEEEproof}

We show the result for
$\lambda_e=\Theta\left((\log n)^{-(2+\rho)}\right)$,
which will imply the desired result (as lowering
the eavesdropper density can not degrade the performance).
Consequently, there exists constants $\underline{\Lambda}$,
$\overline{\Lambda}$, and $n_3$ such that
\begin{equation}
\underline{\Lambda} (\log n)^{-(2+\rho)} \leq \lambda_e
\leq \overline{\Lambda} (\log n)^{-(2+\rho)}, \textrm{ for }
n\geq n_3,
\end{equation}
where $\underline{\Lambda} < \overline{\Lambda}$.

At this point, we can upper bound the fraction of nodes that can not
access to a highway due to an existence of an eavesdropper in their
first secrecy zone. Following the analysis
in Lemma~\ref{thm:Lemma5:AccessRate}, as long as we satisfy
\begin{eqnarray}\label{eq:s11e1}
\lambda_e (f_{l_1})^2 d^2 \to 0, \textrm{ as } n\to\infty,
\end{eqnarray}
almost all the nodes can access to the highways.
To compute the achievable secrecy rate with
Lemma~$1$, we need to satisfy
\begin{eqnarray}\label{eq:s11e2}
\lambda_e (f_{l_k})^2 d^2 \to \infty, \textrm{ as } n\to\infty,
\textrm{ for } k=2,3,\cdots.
\end{eqnarray}
Further, we can show that as long as we satisfy
\begin{eqnarray}\label{eq:s11e3}
\lambda_e d^2 \sum\limits_{k=2}^L (f_{l_k})^2 (f_{l_{k-1}})^{-\alpha}\leq
C, \textrm{ as } n\to\infty,
\end{eqnarray}
for some constant $C$,
the achievable rate $R_{TR}$ in Lemma~\ref{thm:Lemma1:ColludingSecureRate}
scales like $\Omega\left((\log n)^{-2-\alpha}\right)$
as $d=\kappa'' \log n$. Due to
time-division among the legitimate nodes accessing the
highways (there are w.h.p. $O(\log n)$ nodes within small
squares), the secrecy rate per user satisfies
$\Omega\left((\log n)^{-3-\alpha}\right)$.

Here, to satisfy \eqref{eq:s11e1},
\eqref{eq:s11e2}, \eqref{eq:s11e3} with
$d=\kappa'' \log n$,
we choose the secrecy zones as
\begin{equation}
f_{l_k} = (\log n)^{r(\frac{\alpha}{2})^{k-1}},
\end{equation}
with some $r$ satisfying $\frac{\rho}{\alpha} < r < \frac{\rho}{2}$.

\end{IEEEproof}

Note that, Lemma~\ref{thm:Lemma2:MultihopSecurity}
that the per hop security implies the multi-hop security
also holds for the colluding eavesdropper scenario.
That is, the security obtained for configuration $\Ec^*$
for each hop is sufficient to ensure secrecy against
colluding eavesdroppers listening all the hops.
Combining these results with the percolation result given in
Lemma~\ref{thm:Lemma3:Percolation} concludes the proof.

\newpage

\bibliographystyle{IEEEtran}

\newpage

\begin{figure}[t] 
    \centering
    \includegraphics[width=0.6\columnwidth]{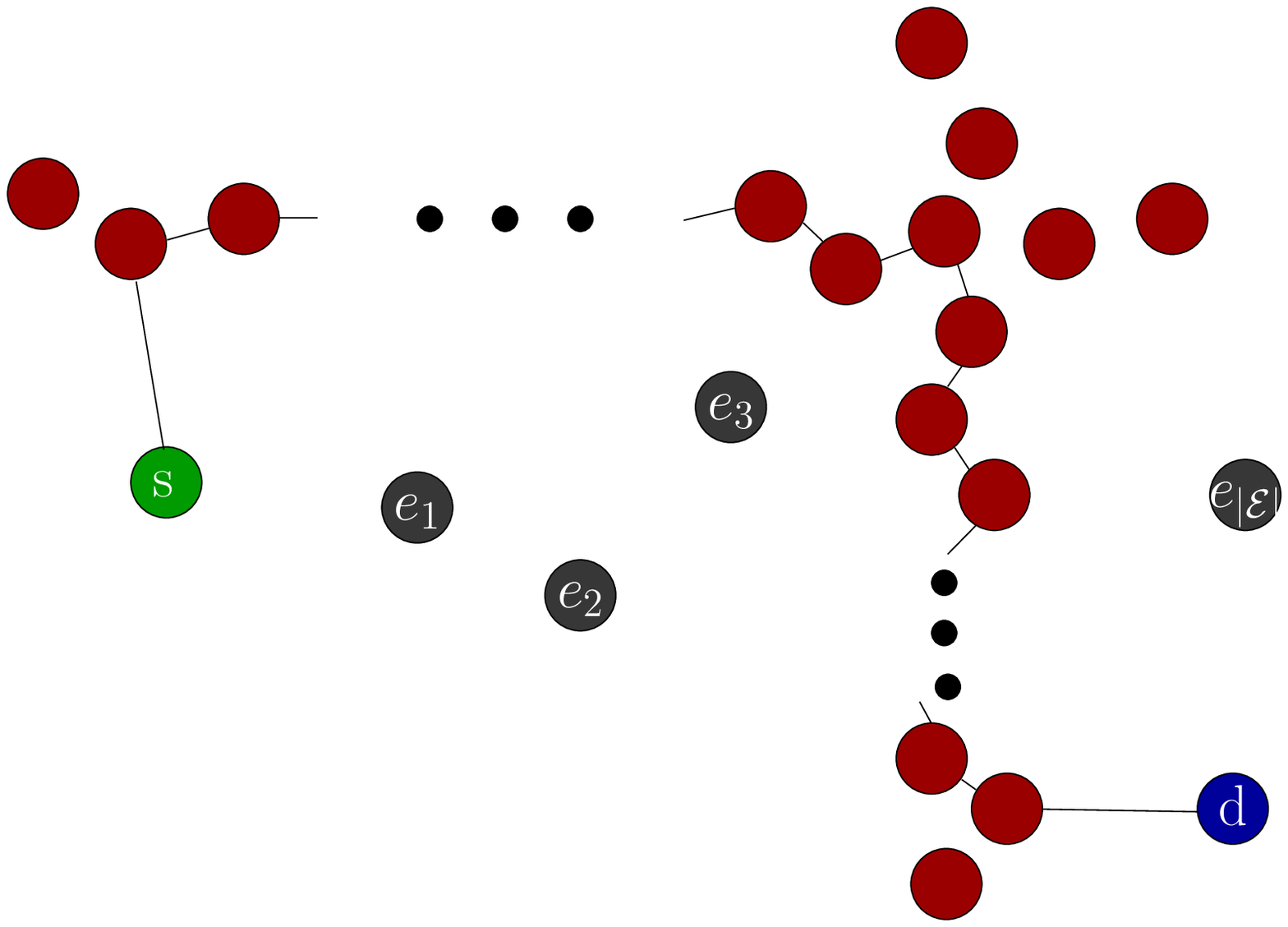}
    \caption{A typical multi-hop route consists of four
    transmission phases:
    $1$) From source node to an entry point on the horizontal highway,
    $2$) Across horizontal highway (message is carried until
    the desired vertical highway member),
    $3$) Across vertical highway (message is carried until
    the exit node), and
    $4$) From the exit node to the destination node.
    }
\end{figure}

\begin{figure}[t] 
    \centering
    \includegraphics[width=0.4\columnwidth]{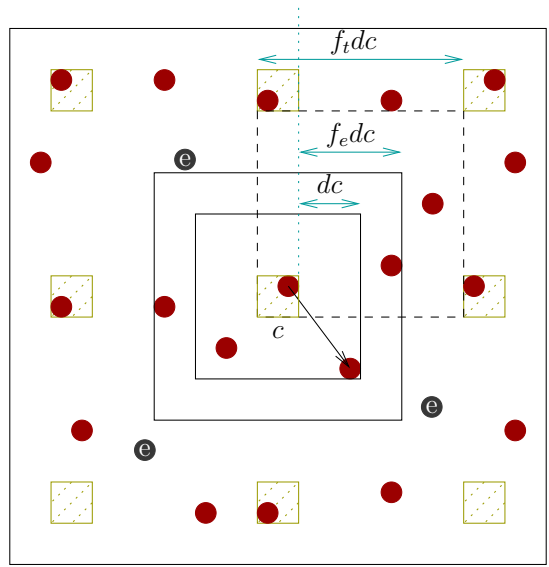}
    \caption{The transmitter located
    at the center of the figure wishes to communicate with a
    receiver that is $d$ squares away. The second square
    surrounding the transmitter is the secrecy zone, which is the
    region of points that are at most $f_e\:d$ squares away from
    the transmitter. Side length of each square is denoted by $c$.
    The time division approach is represented by
    the shaded squares that are allowed for transmission.
    It is evident from the dashed square that the time division
    requires $(f_t\:d)^2$ time slots.}
\end{figure}

\begin{figure}[t] 
    \centering
    \includegraphics[width=0.3\columnwidth]{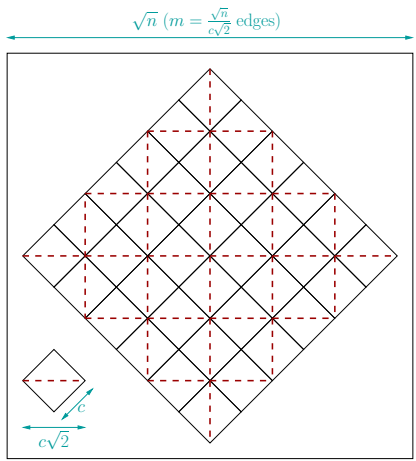}
    \caption{Horizontal and vertical edges in the discrete bond
    percolation model are denoted by dotted lines. A dotted edge
    is open (used for the highway construction) if the corresponding
    square is open. There are $\Theta(n)$ number
    of edges in the random graph.}
\end{figure}

\begin{figure}[t] 
    \centering
    \includegraphics[width=0.36\columnwidth]{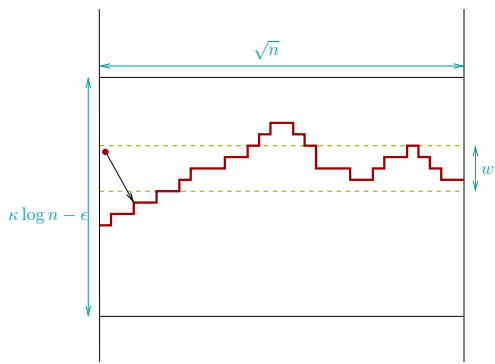}
    \caption{There are $\lceil \delta \log n \rceil$
    number of disjoint highways within each rectangle of
    size $(\kappa \log n - \epsilon) \times \sqrt{n}$.
    The legitimate users in the slab, denoted by dotted lines,
    of the rectangle is served by the highway denoted with
    red bold line.}
\end{figure}

\begin{figure}[t] 
    \centering
    \includegraphics[width=0.4\columnwidth]{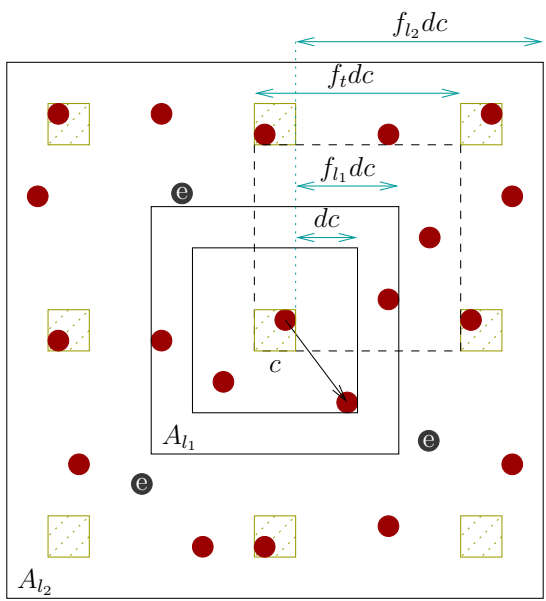}
    \caption{The second square
    surrounding the transmitter is the secrecy zone (zone of level
    $1$), which is the region of points that are
    at most $f_{l_1} d$ squares away from
    the transmitter. The zone of level $k$ is denoted with distance
    $f_{l_k} d c$ and has an area of $A_{l_k}$.}
\end{figure}

\end{document}